\documentclass[review]{elsarticle}

\usepackage{lineno,hyperref}
\modulolinenumbers[5]

\journal{International Journal of Heat and Fluid Flow}

\biboptions{authoryear}

\usepackage{graphicx}
\usepackage{amsmath}
\usepackage{amssymb}

\usepackage{xcolor}	
\definecolor{excel1}{gray}{0}
\definecolor{excel2}{gray}{0.5}
\definecolor{excel3}{gray}{0.8}
\definecolor{excel4}{gray}{0.3}
\definecolor{blueFig6}{rgb}{0,0.4470,0.7410}
\begin{document}

\begin{frontmatter}

\title{Linear control of coherent structures in wall-bounded turbulence at Re$_\tau = 2000$}

\author{Stephan F. Oehler\corref{cor1}} 

\cortext[cor1]{Corresponding author, Email: stephan.friedrich.oehler@gmail.com}

\author{Simon J. Illingworth}%

\address{Department of Mechanical Engineering,
	The University of Melbourne, Victoria 3010, Australia}

%\email{stephan.friedrich.oehler@gmail.com}

%\title{Elsevier \LaTeX\ template\tnoteref{mytitlenote}}
%\tnotetext[mytitlenote]{Fully documented templates are available in the elsarticle package on \href{http://www.ctan.org/tex-archive/macros/latex/contrib/elsarticle}{CTAN}.}
%
%%% Group authors per affiliation:
%\author{Elsevier\fnref{myfootnote}}
%\address{Radarweg 29, Amsterdam}
%\fntext[myfootnote]{Since 1880.}
%
%% or include affiliations in footnotes:
%\author[mymainaddress,mysecondaryaddress]{Elsevier Inc}
%\ead[url]{www.elsevier.com}
%
%\author[mysecondaryaddress]{Global Customer Service\corref{mycorrespondingauthor}}
%\cortext[mycorrespondingauthor]{Corresponding author}
%\ead{support@elsevier.com}
%
%\address[mymainaddress]{1600 John F Kennedy Boulevard, Philadelphia}
%\address[mysecondaryaddress]{360 Park Avenue South, New York}

\begin{abstract}
We consider linear feedback flow control of the largest scales in an incompressible turbulent channel flow at a friction Reynolds number of $Re_\tau = 2000$. A linear model is formed by linearizing the Navier-Stokes equations about the turbulent mean and augmenting it with an eddy viscosity. Velocity perturbations are then generated by stochastically forcing the linear operator. The objective is to reduce the kinetic energy of these velocity perturbations at the largest scales using body forces. It is shown that a control set-up with a well-placed array of sensors and actuators performs comparably to either measuring the flow everywhere (while limiting actuators to a single wall height) or actuating the flow everywhere (while limiting sensors to a single  wall height). %This idealized configuration, therefore, can provide insight into how specific scales of turbulence are most effectively measured and actuated at low computational cost. 
In this way, we gain insight (at low computational cost) into how the very large scales of turbulence are most effectively estimated and controlled.
\end{abstract}

\begin{keyword}
Turbulence\sep Channel flow\sep Control \sep Linear model \sep
Coherent structures %\sep 
\end{keyword}

\end{frontmatter}

\section{Introduction}
\label{J2:Intro}
A growing number of studies have successfully utilized linear models for estimation  \citep[e.g.][]{Chevalier2006,jones2011flow,illingworth2018estimating,oehler2018linear,sasaki2019transfer} and control \citep[e.g][]{cortelezzi1998robust,moarref2012model,luhar2014opposition} of wall-bounded turbulent flows. The work of \citet{luhar2014opposition}, in particular, suggests that linear models can qualitatively predict the effect of control on individual scales and also determine at which location they can best be measured. Linear model-based designs are an appealing alternative to direct numerical simulation (DNS) based designs since the cost is several orders of magnitude smaller. 	
One reason for the success of linear models is that linear mechanisms play an important role in the sustenance of turbulence \citep{schoppa2002coherent,kim2011physics}. In the linearized Navier-Stokes (LNS) equations, where the flow is linearized around the turbulent mean, these linear mechanisms result in large transient growth that is due to the non-normality of the LNS operator \citep{trefethen1993hydrodynamic}.
In particular, it was shown that the LNS operator could predict the typical widths of near-wall streaks and large-scale structures in the outer layer \citep{DelAlamo2006,Pujals2009,hwang2010linear}.

Linear mechanisms play a major role in the formation and maintenance of large-scale structures in turbulent wall-bounded flows.~These large-scale structures contribute significantly to the turbulent kinetic energy and Reynolds stresses (in the outer region), and there is evidence that they affect the small scales near the wall \citep{hutchins2007large,mathis2009large,marusic2010high,marusic2010predictive,duvvuri2015triadic}.
Hence, the control of these structures is crucial for any efforts to control wall-bounded flows (see \citet{Kim2007} for a review). 
It was shown that linear estimation, which is closely related to linear control, performs best for those structures that have the greatest potential for transient growth \citep{DelAlamo2006,Pujals2009}, are the most amplified in stochastically and harmonically forced settings \citep{hwang2010linear} and are coherent over large wall-normal distances \citep{Madhusudanan2019SLSE}. These observations presumably also apply to linear control, which would simplify the controller design process.

This work studies linear feedback control of the largest structures in a turbulent channel flow at a relatively high Reynolds number of Re$_\tau=2000$. It is in part motivated by experimental work that has achieved a reduction in skin-friction drag through real-time control of large-scale structures \citep{abbassi2017skin}. The focus of this study is on the sensors and actuators for linear  feedback-control. Specifically, we compare control performance when measuring or actuating the full channel (i.e. an ideal set-up) to control performance when measuring or actuating at only one specific wall height (which is a more realistic set-up in a practical application, e.g. hot-wire sensors and synthetic jet actuators). Consequently, it is possible to compare the ideal setting to what is achievable in a laboratory environment.

When considering these control set-ups, we will focus on finding the best control performance possible.
Therefore, we (i) assume that sensor noise is insignificant, (ii) remove almost all energy limitations imposed on the actuators, and (iii) ignore the effect of transients. In this way, it is possible to show whether a setup is worth considering in the first place as even the best results might not be sufficient.

Rather than testing various control configurations through the use of DNS, the flow is modeled using the LNS operator for perturbations about the mean flow (\S \ref{J2:sec:LM}). An eddy viscosity is included in the operator to model the effect of the incoherent scales. This uncontrolled linear model (LM) of the flow is validated by comparing it to DNS in \S \ref{J2:sec:validation} before we introduce three specific control set-ups in \S \ref{J2:sec:LC} and analyze their performance in \S \ref{J2:sec:results}. Finally, we conclude the study in \S \ref{J2:sec:discussion}.

\section[The linear model]{The linear model}

\label{J2:sec:LM}

A statistically steady incompressible turbulent channel flow at a friction Reynolds number Re$_\tau = u_\tau h / \nu = 2000$ is considered, where $\nu$ is the kinematic viscosity, $h$ the channel half-height, $u_\tau = \sqrt{\tau_w  / \rho}$ the friction velocity, $\tau_w$ the wall shear stress, and $\rho$ the density. 
Streamwise, spanwise, and wall-normal spatial coordinates  are denoted by $[x,y,z]$ and the corresponding velocities by $[u,v,w]$. We assume zero initial conditions and apply no-slip boundary conditions. Spatial variables are normalized by $h$, wavenumbers by $1/h$, velocities by the friction velocity $u_\tau$, time by $h/u_\tau$ and pressure $p$ by $\rho u_\tau^2$. This non-dimensionalization sets the channel half-height to $h = 1$ such that $z \in [0,2h]$.

Following \cite{Reynolds1972}, we triple decompose the overall velocity field $\tilde{\boldsymbol{u}}$ of the turbulent channel into
\begin{align}
\tilde{\boldsymbol{u}} = \boldsymbol{ U } + \boldsymbol{u} + \boldsymbol{u^{'}},
\end{align}
where $\boldsymbol U$ represents the turbulent mean flow, $\boldsymbol{u}$ large scale organised motion (or waves) and $\boldsymbol{u^{'}}$ small scale turbulent fluctuations.
Taking the incompressible Navier-Stokes equations we form a linear operator for the perturbations $\boldsymbol{u} = \left[u,v,w\right]$ about the turbulent mean flow $\boldsymbol{U} = \left[U(z),0,0\right]$, where the non-linear term $\boldsymbol{ {d} } = - (\boldsymbol{ u} \cdot \nabla)\boldsymbol{u} + \overline{(\boldsymbol{u} \cdot \nabla) \boldsymbol{u}}$ is treated as stochastic forcing and ($\overline{\cdot}$) is the time-averaged mean. An eddy viscosity $\nu_T(z)$, which accounts for the average dissipative effect of the stresses created by the small scale turbulent fluctuations ($\boldsymbol{u^{'}}$), is introduced to represent the influence of incoherent motions \citep{Reynolds1972,DelAlamo2006,Pujals2009,hwang2010linear,eitel2015hairpin,hwang2016mesolayer,hwang2017mesolayer}:
\begin{align}
\frac{\partial \boldsymbol{{u}}}{ {\partial t}} + \left(\boldsymbol{ U } \cdot \nabla\right) \boldsymbol{ u } - \left(\boldsymbol{ u } \cdot \nabla\right) \boldsymbol{ U} = - \nabla p + \nabla \cdot \left[\frac{\nu_T}{\nu} (\nabla \boldsymbol{ u} + \nabla \boldsymbol{ u}^T)\right] + \boldsymbol{d},
\label{J2:eqn:NVSmp} \hspace{10mm}
\nabla \cdot \boldsymbol{u}  = 0.
\end{align}

An analytical fit is used \citep{cess1958survey} for the eddy viscosity profile $\nu_T$ as in several previous studies \citep{Pujals2009,DelAlamo2006,moarref2012model,illingworth2018estimating}: \begin{align}
\nu_T(z) = \frac{\nu}{2}\left\{1 + \frac{\kappa_1^2 \text{Re}_\tau^2 }{9}(2z - z^2)^2(3 - 4z + 2z^2    )^2
\times \left[1 - \exp\left(\frac{-\text{Re}_\tau z}{\kappa_2}\right)\right]^2 \right\}^{{1}/{2}} + \frac{\nu}{2}.
\label{J2:eqn:eddyprofile}
\end{align}
Integrating Re$_\tau(1 - z) \nu / \nu_T(z)$ provides the mean velocity profile $U(z)$. The constants $\kappa_1 = 0.426$ and $\kappa_2 = 25.4$ give the best fit to the mean velocity profile of a DNS simulation at Re$_\tau = 2003$ \citep{Hoyas2006,DelAlamo2006}. 
Controlling perturbations in the flow will alter the mean velocity profile and with it the linear model itself (which is formed about the mean). The controller, therefore, needs to be robust to account for any changes in the mean flow. It would be interesting to study robustness, but this is beyond the scope of this study. 

We need to express the flow in state-space form to access standard tools from dynamics and control. 
To do so, we first take Fourier transforms in the homogeneous streamwise and spanwise directions to express the flow in the Orr-Sommerfeld Squire form and then discretize in the wall-normal direction using Chebyshev collocation of order $N = 200$. Convergence has been checked for all control set-ups by doubling the number of grid points; it was shown that the tested results change by less than $0.6\%$ (see \ref{J2:sec:converg}). Finally, we express the Orr-Sommerfeld Squire equations as a linear state-space model:

\begin{minipage}{.48\textwidth}
	\vspace{1 mm}
	\centering
	\includegraphics[width=1\linewidth]{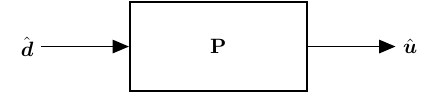}	
	\label{blockP1ss}
\end{minipage}
\begin{minipage}[lo]{.48\textwidth}
	\begin{subequations}
		\vspace{-1 mm}
		\begin{align}
		\dot{\hat{\boldsymbol{q}}}(t) &= \mathbf{A} \hat{\boldsymbol{q}}(t) + \mathbf{B}_d\hat{\boldsymbol{d}}(t),\label{J2:eqn:ssp1}
		\\
		\hat{\boldsymbol{u}}(t) &= \mathbf{{C}} {\hat{\boldsymbol{q}}}(t),
		\label{J2:eqn:ssp2}
		\end{align}
		\label{J2:eqn:ssp}
	\end{subequations}
\end{minipage}
where $\hat{\boldsymbol{q}} = \left[\hat{\boldsymbol{w}},\hat{\boldsymbol{\eta}}\right]^T$ represents the states of the system (wall-normal velocity and wall-normal vorticity), $\hat{\boldsymbol{d}} = \left[\hat{{d}}_x,\hat{{d}}_y,\hat{{d}}_z\right]^T$ all non-linearities and $\hat{\boldsymbol{u}} = [\hat{u},\hat{v},\hat{w}]$ the velocities ( ( $\hat{}$ ) denotes signals in Fourier space). We treat $\hat{\boldsymbol{d}} $ as stochastic forcing that is white in wavenumber space and time \citep{jovanovic2005componentwise}. Therefore, we account for the non-linearities by treating them as a source of intrinsic forcing to the LNS operator \citep{mckeon2010critical}. We set $\mathbf{B}_d = \mathbf M^{-1/2} \mathbf B$ to achieve grid-independence, where $\mathbf M$ is an integration matrix corresponding to Clenshaw--Curtis quadrature \citep{trefethen2000spectral}, and we choose $\mathbf C$ in equation \eqref{J2:eqn:ssp2} such that $\hat{\boldsymbol{u}}$ corresponds to the velocity field over one channel-half ($0 \leq z\leq h$). 
The matrices $\mathbf A$, $\mathbf B$, and $\mathbf C$ are:
\begin{align}
\mathbf A &=
\begin{bmatrix}
\Delta^{-1}\mathcal{L}_{OS} & 0 \\ -\mathrm{i}k_y U^\prime & \mathcal{L}_{SQ}
\end{bmatrix}, 
\\
\mathbf B &=  \begin{bmatrix}
-\mathrm{i}k_x \Delta^{-1} \mathcal{D}  && -\mathrm{i}k_y \Delta^{-1} \mathcal{D} && -k^2 \Delta^{-1} \\ \mathrm{i}k_y  && -\mathrm{i}k_x && 0
\end{bmatrix},  
\\
\mathbf C
&=
\frac{1}{k^2}
\begin{bmatrix}
\mathrm{i}k_x  \mathcal{D} & -\mathrm{i}k_y
\\
\mathrm{i}k_y  \mathcal{D} & \mathrm{i}k_x
\\
k^2 & 0 
\end{bmatrix}, 
\label{J2:eqn:ss2}
\end{align}
where $\mathcal{L}_{OS}$ and $\mathcal{L}_{SQ}$ are the Orr-Sommerfeld and Squire operators for the eddy viscosity enhanced LNS equations \citep{betchov1966spatial,Pujals2009}: 
\begin{align}
\mathcal{L}_{OS} &= \mathrm{i} k_x (U^{\prime\prime} - U \Delta)  + \nu_T \Delta^2 + 2 \nu_T^\prime \mathcal{D} \Delta + \nu_T^{\prime\prime}\left(\mathcal{D}^2 + k^2\right), 
\\
\mathcal{L}_{SQ} &= -\mathrm{i}k_x U + \nu_T \Delta + \nu_T^\prime \mathcal{D}.
\end{align}
Here $\mathcal{D} = \frac{\partial}{\partial z},$ $()^\prime = \frac{\partial}{\partial z}()$, $k^2 = k_x^2 + k_y^2$, and $ \Delta = \mathcal{D}^2 - k^2$. The boundary conditions are: $\hat{\boldsymbol{w}}_{wall}(t) = \hat{\boldsymbol{w}}_{wall}^\prime(t) = \hat{\boldsymbol{\eta}}_{wall}(t) = 0$. (See \ref{J2:apdx:1} for more information.) 
By taking Laplace transforms of equation \eqref{J2:eqn:ssp} we obtain a transfer function $\mathbf P$ that relates the input $\hat{\boldsymbol{d}}$ to the output $\hat{\boldsymbol{u}}$:
\begin{subequations}
	\begin{align}
	\hat{\boldsymbol{u}}(s) &= \mathbf{P}(s) \hat{\boldsymbol{d}}(s),
	\\
	\mathbf{P}(s) &= \mathbf{C}\left(s\mathbf{I} - \mathbf{A}\right)^{-1} \mathbf B_d,
	\end{align}
	\label{J2:eqn:ssP}
\end{subequations}
where $s$ is the Laplace variable. By setting $s = \mathrm j\omega$ the frequency response (i.e. the resolvent) is obtained.

%------------------------------------------------%

We quantify the energy of the flow by employing the the square of the $\mathcal{H}_2$-norm (\ref{J2:APPDIX:h2norms}) of $\hat{\boldsymbol u}$:
\begin{subequations}
\begin{align}
\|\hat{\boldsymbol u}\|_2^2  \equiv&  { \dfrac{1}{2\pi} \int_{-\infty}^{\infty} \text{trace}\left[\mathbf{P}^*(\mathrm{j} \omega) \mathbf{M} \mathbf{P} (\mathrm{j} \omega)\right] d \omega }\\ \equiv&  \mathbb{E} \left\{  \lim\limits_{T \rightarrow \infty} \frac{1}{T} \int_{0}^{T} \int_{0}^{h} {\hat{\boldsymbol{u}}^*(z,t)}\hat{\boldsymbol{u}}(z,t)dz dt\right\},
\label{J2:Energynorm_u}	
\end{align}	
\label{J2:eqn:LM:gam}
\end{subequations}
where $\mathbb{E}$ is the expected value and $()^*$ is the complex conjugate transpose. 
In the Laplace domain, the $\mathcal{H}_2$-norm of $\mathbf P(s = \mathrm j \omega)$ can be seen as the average gain between the input $ \hat{\boldsymbol d}(s = \mathrm{j}\omega)$ and the output $\hat{\boldsymbol{u}}(s = \mathrm{j}\omega)$ over all frequencies and all directions.

In \S \ref{J2:sec:results:large}--\S\ref{J2:sec:forcing}, we choose to focus on the streamwise and spanwise wavenumber pairs that are most amplified (we select them to be $|k_x| \leq 0.5$ and $ |k_y| \leq 6$).	
In particular, we are interested in
the energy as a function of wall height for this range of wavenumbers. First, we obtain the square of the $\mathcal{H}_2$-norm for individual wavenumber pairs:
\begin{subequations}
\begin{align}
\|\hat{\boldsymbol u}(z)\|_2^2  \equiv&  { \dfrac{1}{2\pi} \int_{-\infty}^{\infty} \text{trace}\left[\mathbf{P}^*(\mathrm{j} \omega) \mathbf{P} (\mathrm{j} \omega)\right] d \omega }\\ \equiv&   \mathbb{E} \left\{  \lim\limits_{T \rightarrow \infty} \frac{1}{T} \int_{0}^{T} {\hat{\boldsymbol{u}}^*(z,t)}\hat{\boldsymbol{u}}(z,t) dt\right\}.
\label{J2:Energynorm_uz}	
\end{align}
\end{subequations}
By summing $\|\hat{\boldsymbol u}(z)\|_2^2$: 
\begin{align}
\|{\boldsymbol u}(z)\|_2^2 =  \sum_{i\in k_x, j \in k_y} \|\hat{\boldsymbol u}(i,j,z)\|_2^2,
\label{J2:eqn:sum_rms}
\end{align}
the square of the $\mathcal{H}_2$-norm for the chosen set of wavenumbers pairs as a function of wall height is obtained. (Please note that this summation is only valid if the wavenumbers are uniformly spaced.)

\section{Validation of the linear model with DNS}
\label{J2:sec:validation}
To validate the linear model, we employ a direct numerical simulation (DNS) dataset provided by the Polytechnic University of Madrid \citep{Hoyas2006,encinar2018second}. We will look at (i) the DNS model itself,
 (ii) the flow's energy as a function of wavenumber ($k_x$ and $k_y$), (iii) the flow's energy as a function of wall height and (iv) a snapshot of streamwise velocity perturbations.

\subsection{Direct Numerical Simulation (DNS)}

The homogeneous streamwise and spanwise directions (extending $8 \pi \times 3\pi$) of the turbulent channel flow (in DNS) are discretised by Fourier expansion (with a streamwise resolution of $\Delta k_x = 1/4$ and a spanwise resolution of $\Delta k_y = 2/3$), and the wall-normal direction is discretised using a compact difference scheme of 7th order.  The data is real-valued in physical space, and therefore, the coefficients for modes ($k_x,+k_y$) are the same as those for ($k_x,-k_y$).  We consider data for every $\delta t = 0.0111$ terminated at $t_{max} = 12.7$. A total of $t_{max}U_c / (8 \pi) = 12.3$ channel flow-throughs ensures that any transients in the estimators and controllers are negligible (where $U_c$ is the mean velocity at the channel centre). The largest temporal frequency is approximated using Taylor's hypothesis: $\omega_{max} = \textrm{max}(|k_x|) U_c = 12.2$, where $U_c$ is the velocity at the channel centre, and max$(|k_x|)$ the largest streamwise wavenumber considered. Therefore, we have $2\pi/ (\omega_{max} \Delta t) = 46.5$ samples per period for the highest frequency, which fulfils the Nyquist criterion. 

To quantify the energy of the DNS data, we compute the square of the $L_2$-norm by integrating $\hat{\boldsymbol u}^* \hat{\boldsymbol u}$ in time and space: 
\begin{align}
\|\hat{\boldsymbol u}\|_2^2 &= \int_{0}^{t_{max}} \int_{0}^{h} {\hat{\boldsymbol{u}}^*(z,t)}\hat{\boldsymbol{u}}(z,t)dz dt,
\label{J2:eqn:DNS:gam}
\end{align}
which is the square of the $L_2$-norm for one channel half;
and by integrating  $\hat{\boldsymbol u}^* \hat{\boldsymbol u}$ in time only:
\begin{align}
\|\hat{\boldsymbol u}(z)\|_2^2 &= \int_{0}^{t_{max}}  {\hat{\boldsymbol{u}}^*(z,t)}\hat{\boldsymbol{u}}(z,t) dt,
\label{J2:eqn:DNS:rms}
\end{align}
which is the square of the $L_2$-norm at individual wall heights.

\subsection{Energy as a function of wavenumber}

\begin{figure}
	\vspace{2mm}
	\centering
	\includegraphics[width=0.99\textwidth,trim={0cm 0cm 7.25cm 0cm},clip]{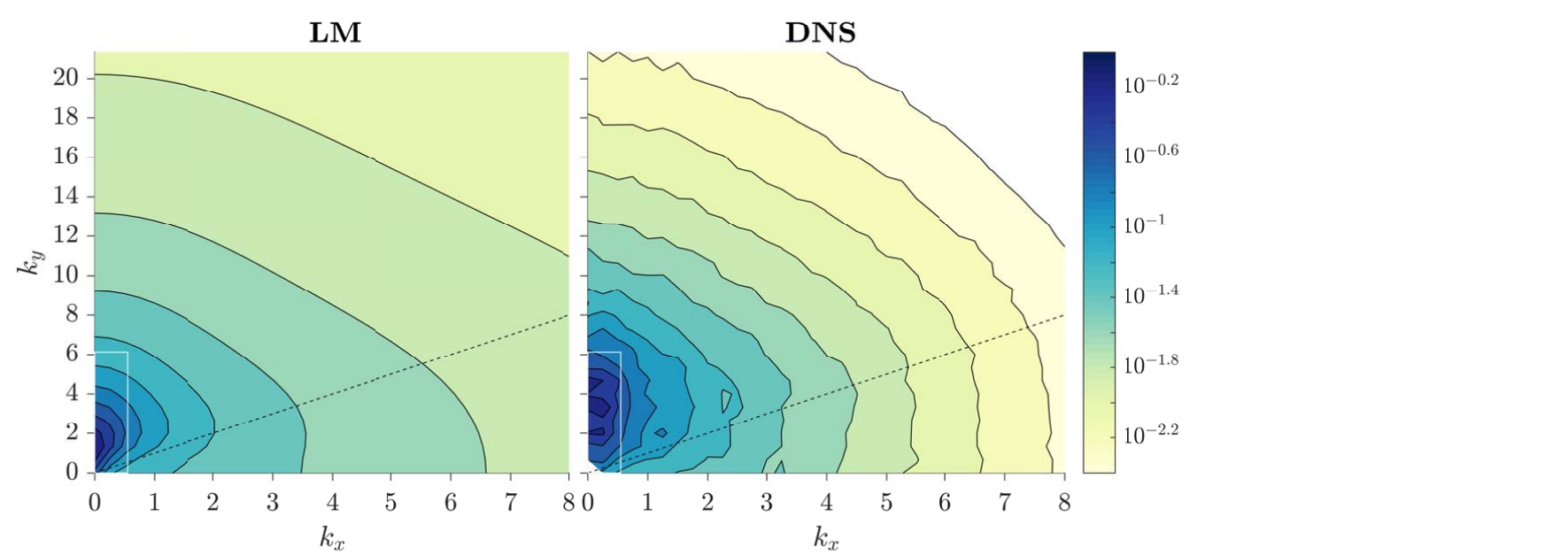}
	\caption[The energy of the uncontrolled flow as a function of $k_x$ and $k_y$]{ The energy of the uncontrolled flow for the LM and for DNS as a function of streamwise $k_x$ and spanwise $k_y$ wavenumber represented by contour levels from $10^{-2.4}$ (yellow) to $1$ (blue).~The results are normalized to $1$ and presented on a logarithmic scale.~Also denoted on the figure are the wavenumber pairs where $k_x = k_y(--)$ and the range of wavenumbers considered for control throughout the rest of this study$(\square)$.
	} 
	\label{J2:fig:kxkyrange_u}
\end{figure}

The flow's energy $\|\hat{\boldsymbol u}\|_2^2$ as a function of wavenumber ($k_x$ and $k_y$) is calculated using equation \eqref{J2:eqn:LM:gam} for the LM and \eqref{J2:eqn:DNS:gam} for DNS and displayed in Figure \ref{J2:fig:kxkyrange_u}, which shows $ \|\hat{\boldsymbol u}\|_2^2$ for a range of $k_x$ and $k_y$. The results are normalised to $1$ (the maximum value in each plot) and presented on a logarithmic scale. 

We can see similarities between DNS and the LM, especially for the selected set of $|k_x| \leq 0.5$ and $ |k_y| \leq 6$. However, the energy degrades more slowly with increasing $k_x$ and $k_y$ in the LM relative to DNS. In addition, the peaks of $\|\hat{\boldsymbol u}\|_2^2$ do no match. They are located at $k_x = 0$ and $k_y = 4/3$ for the LM and $k_x = 0.25$ and $k_y = 10/3$ for DNS. For the purposes of this study, this match is sufficient. Refer to \citet{hwang2010linear,hwang2010amplification} for a detailed analysis on the energy amplification of the turbulent channel flow.

\subsection{Energy as a function of wall height}

\begin{figure}
	\vspace{2mm}
	\centering
	\includegraphics[width=0.7\textwidth,trim={0cm 0cm 0cm 0cm},clip]{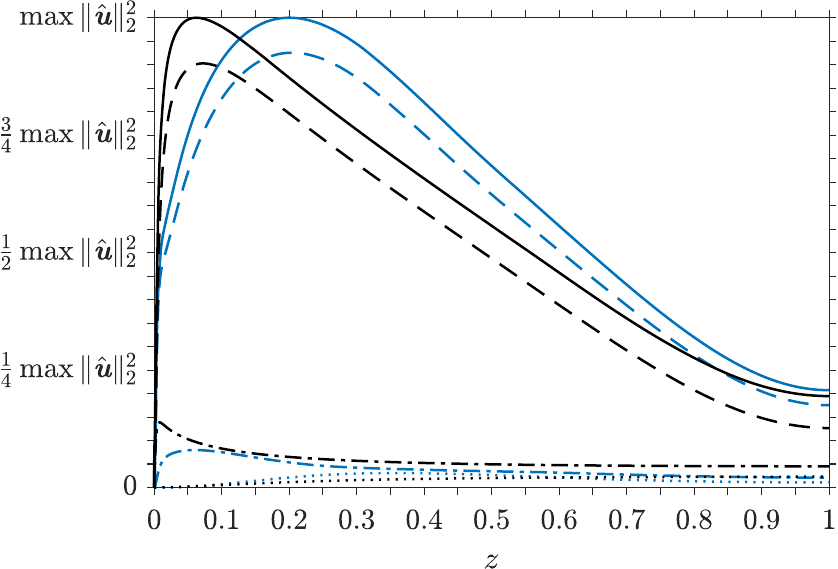}
	\caption[ The energy of the uncontrolled flow as a function of wall height]{ The energy of the uncontrolled flow as a function of wall height for $|k_x| \leq 0.5$ and $ |k_y| \leq 6$. Results are provided for the LM and {\color{blueFig6}DNS}. The energy is shown for all directions $\|\hat{\boldsymbol u}\|_2^2$(---), the streamwise direction $\|\hat{ u}\|_2^2$($--$), the spanwise direction $\|\hat{v}\|_2^2$($\cdot - \cdot$) and the wall-normal direction $\|\hat{ w}\|_2^2$($\cdots$). }
	\label{J2:fig:umaxwallheight}
\end{figure}

We now focus on the set of wavenumbers: $|k_x| \leq 0.5$ and $ |k_y| \leq 6$. For this range, we employ  equations \eqref{J2:eqn:sum_rms} and \eqref{J2:eqn:DNS:rms} to calculate the energy as a function of wall height ($\|{\boldsymbol u}(z)\|_2^2$), which we show in figure \ref{J2:fig:umaxwallheight}. As before, all results are normalised to the maximum energy value for the LM and DNS respectively.

We see that the flow is most energetic at $z = 0.06$ for the LM and $x = 0.20$ for DNS. We also observe that for DNS, the energy is more evenly distributed throughout the channel. Looking at the individual flow directions, we see that (i) the streamwise velocity fluctuations are the most energetic, (ii) the spanwise velocity fluctuations behave differently at the wall, and (iii) the wall-normal velocity fluctuations are the least energetic.

\subsection{Snapshot of streamwise velocity perturbations}

\begin{figure}
	\vspace{2mm}
	\centering
	\includegraphics[width=0.8\textwidth,trim={0cm 7cm 0cm 0cm},clip]{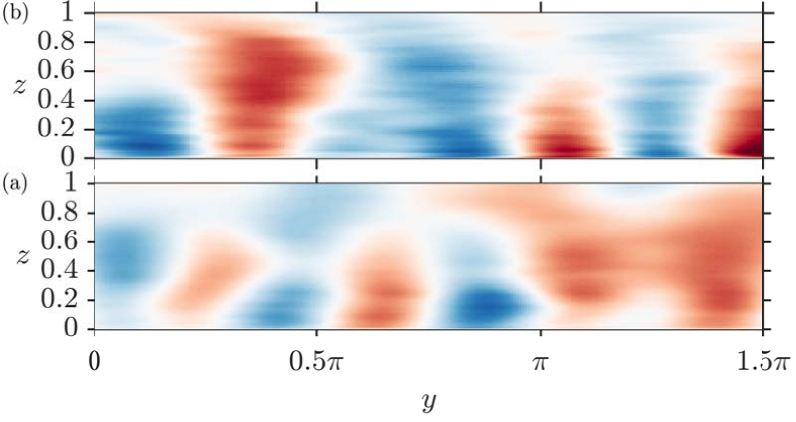}
	\caption[Streamwise velocity perturbations at $x = 3\pi/2$]{Streamwise velocity perturbations at $x = 3\pi/2$ for $|k_x| \leq 0.5$ and $ |k_y| \leq 6$. Results are shown for (a) LM and (b) DNS using sixty-five contour levels from $-\|u\|_{max}$ (blue) to $\|u\|_{max}$ (red). (The snapshots are taken from figures 4(a) and 2(a) in \cite{oehler2018linearchannell}.)}
	\label{afmc:fig:P}
\end{figure}

Finally, in Figure \ref{afmc:fig:P}((a) LM and (b) DNS), we show the streamwise velocity perturbations at $x = 3\pi/2$ in a spanwise wall-normal ($y-z$) plane at an instant in time for $|k_x| \leq 0.5$ and $ |k_y| \leq 6$.
In both cases, we observe large structures that are strongest near the wall and that reduce in strength towards the channel centre, in agreement with figure \ref{J2:fig:umaxwallheight}.

\section[The control set-up]{The control set-up}

\label{J2:sec:LC}

So far, we have introduced the eddy-viscosity-enhanced Orr-Sommerfeld and Squire operators that are linearized about the mean velocity profile of a turbulent channel flow. We stochastically force the linear operator to generate velocity perturbations that we now want to control. To do so, we include three new signals ($\hat{\boldsymbol m}$, $\hat{\boldsymbol f}$ and $\hat{\boldsymbol z}$) into the state-space model (equation \eqref{J2:eqn:ssp}):

\begin{minipage}{.4\textwidth}
	\vspace{5 mm}
	\centering
	\includegraphics[width=1\linewidth]{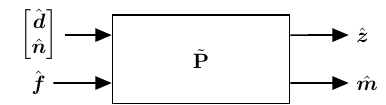}
	\label{fig:blockgen}
\end{minipage}	
\begin{minipage}{.55\textwidth}
	\vspace{-7mm}
	\begin{subequations}
		\label{eqn:Ptilde}
		\begin{align}
		{\dot{\hat{\boldsymbol q}}}(t) &= \mathbf{A} \hat{\boldsymbol q}(t) + \mathbf{B}_d \hat{\boldsymbol d}(t) + \mathbf{B}_f \hat{\boldsymbol f}(t),   \label{eqn:GLE:ss2}	
		\\	\hat{\boldsymbol z}(t) &= \mathbf{C}_z \hat{\boldsymbol q}(t) + 	\alpha   \hat{\boldsymbol f}(t), \label{ss22}
		\\
		\hat{\boldsymbol{m}}(t) &=
		{\mathbf C}_m 
		\hat{\boldsymbol q}(t) + 	\hat{\boldsymbol n}(t)\label{ss21}.
		\end{align}
	\end{subequations} 
\end{minipage}

The first new signal $\hat{\boldsymbol m}$ represents time-resolved velocity measurements from sensors $\mathbf C_m \hat{\boldsymbol q}$ (e.g.~hotwires)  that are contaminated by sensor noise $\hat{\boldsymbol n}$. We treat $\hat{\boldsymbol n}$ as unknown and white in time with a covariance $\mathbb{E}(\hat{\boldsymbol{n}} \hat{\boldsymbol{n}}^*) = n = 10^{-4}$. (The sensor
noise is not correlated between different
wavenumbers.)

The second new signal $\hat{\boldsymbol{f}}$ represents time-resolved body forces applied by actuators (e.g.~synthetic jets $\mathbf{B}_f \hat{\boldsymbol f}$ \citep{cattafesta2011actuators}). 

The third new signal $\hat{\boldsymbol z}$ (not to be confused with wall-normal variable $z$) represents the quantity to be minimized by control, and is derived from a cost function (see \ref{J2:APPDIX:cost}). We define $\hat{\boldsymbol z}$ to minimize the energy of the entire flow field $\boldsymbol{C}_z \hat{\boldsymbol{q}}$ while also keeping the actuation force $\alpha\hat{\boldsymbol{f}}$ small
(where $\alpha$ is a penalization on $\hat{\boldsymbol{f}}$). 
Minimizing the energy of the entire flow-field $\boldsymbol{C}_z \hat{\boldsymbol{q}}$ lets the control design process decide which perturbations to target for the best results. This is in contrast to opposition control, for example, which focuses on wall-normal velocity perturbations to eliminate streamwise streaks \citep{luhar2014opposition}. 
We set the penalization to be insignificant ($\alpha = 10^{-4}$), because we want the results to be insensitive to the choice of $\alpha$. (We cannot set $\alpha$ to zero as this would result in a poorly posed system.) Increasing $\alpha$ will gradually reduce the control performance and energy consumption of the actuator. (See \ref{J2:sec:noiseenergy} for more information.)

\subsection{Sensor and actuator design}
\label{J2:apdx:2}
\label{J2:adx:sensact}

The measurement signal is defined as:
\begin{align}
\hat{\boldsymbol{m}} = \mathbf{C}_m\hat{\boldsymbol{q}} + \hat{\boldsymbol{n}} = \mathbf{C}_y\mathbf{C}\hat{\boldsymbol{q}} + \hat{\boldsymbol{n}} = \mathbf{C}_y\hat{\boldsymbol{u}} + \hat{\boldsymbol{n}},
\end{align}
where $\hat{\boldsymbol{n}}$ is the sensor noise and $\mathbf C_y$ represents the sensor matrix. 
We treat $\hat{\boldsymbol n}$ as an unknown forcing that is white in time, and we set the covariance $\mathbb{E}( \hat{\boldsymbol n}\hat{\boldsymbol n}^*) = n\mathbf{I} = (10^{-4})\mathbf{I} = \mathbf{V}^{1/2}$ such that the sensor noise is negligible but the system is well-posed. The sensor matrix $\mathbf C_y$ is defined as:
\begin{align}
\mathbf{C}_y = 		\begin{bmatrix}
\mathbf g(z_s) & 0 & 0 \\ 
0 & \mathbf g(z_s) & 0 \\
0 & 0 & \mathbf g(z_s)
\end{bmatrix},
\label{J2:eqn:C2:bary}
\end{align}
where
\begin{align}
\mathbf g(z_s) = \exp \left\{ - \left(\frac{ z - z_{s}}{\sigma_s} \right)^2 \right\}^T \mathbf M
\end{align}
is a Gaussian function, ${z} = \left[z_1,z_2 \cdots z_{N_{out}+1}\right]^T$ are Chebyshev grid points (\ref{J2:apdx:1}), $z_s$ is the sensor plane location and $\sigma_s$ defines the width of the Gaussian. We set $\sigma_s = 0.02$, which is equivalent to a $90$\% wall-normal width of $0.06$. There is one Gaussian function for each of the three flow components (streamwise, spanwise and wall-normal).

The actuator force is $\hat{\boldsymbol{f}}$, and it is applied at a single wall-normal location $(z_a)$ via the matrix $\mathbf{B}_f$ (equation \eqref{eqn:GLE:ss2}):
\begin{align}
\mathbf B_f \hat{\boldsymbol{f}} = 
\mathbf{B} \begin{bmatrix}
\mathbf{h}(z_a) & 0 & 0 \\
0 & \mathbf{h}(z_a) & 0 \\
0 & 0 & \mathbf{h}(z_a)
\end{bmatrix}
\begin{bmatrix}
\hat f_x \\ \hat f_y \\ \hat f_z
\end{bmatrix},
\end{align}
where
\begin{align}
\mathbf h(z_a) = \exp \left\{ - \left(\frac{ z - z_{a}}{\sigma_a} \right)^2 \right\}
\end{align}
is a Gaussian function, $ z = \left[z_1,z_2 \cdots z_{N_{c}+1}\right]^T$ are Chebyshev grid points (\ref{J2:apdx:1}), $z_a$ is the actuator plane and $\sigma_a$ defines the width of the Gaussian.
We set $\sigma_a = 0.02$, which 
is equivalent to a $90$\% wall-normal width of $0.06$. There is one Gaussian function for each flow direction.

The main advantage of using a Gaussian shape is that it approximates the finite thickness of sensors and actuators in physical space. Implementation of this study's sensor setups in physical space would result in a plane of sensors placed at $z_s$ (one sensor for each wavenumber pair considered). There would be evenly spaced arrays of sensors in the streamwise and spanwise direction and the spacing would be determined by $\Delta k_x$ and $\Delta k_y$. Despite being localised at a single wall height, the sensors must cover the whole plane in physical space to act on a particular wavenumber. The same principle would apply to actuators.

However, in Fourier space, we can have a single sensor and a single actuator for each pair of ($k_x,k_y$). Therefore, when we mention single sensors and single actuators, we refer to the placement of the sensors and actuators for a specific wavenumber pair in the context of control design. Once all (single sensor and single actuator) controllers have been designed, it is possible to convert them into physical-space convolution kernels that describe the control rules \citep{bewley1998optimal,hogberg2003linear}.

\subsection{The three control problems}

\begin{figure}
	\begin{minipage}{1\textwidth}
		\centering
		\includegraphics[width=1\textwidth,trim={3cm 13.5cm 22.5cm 3cm},clip]{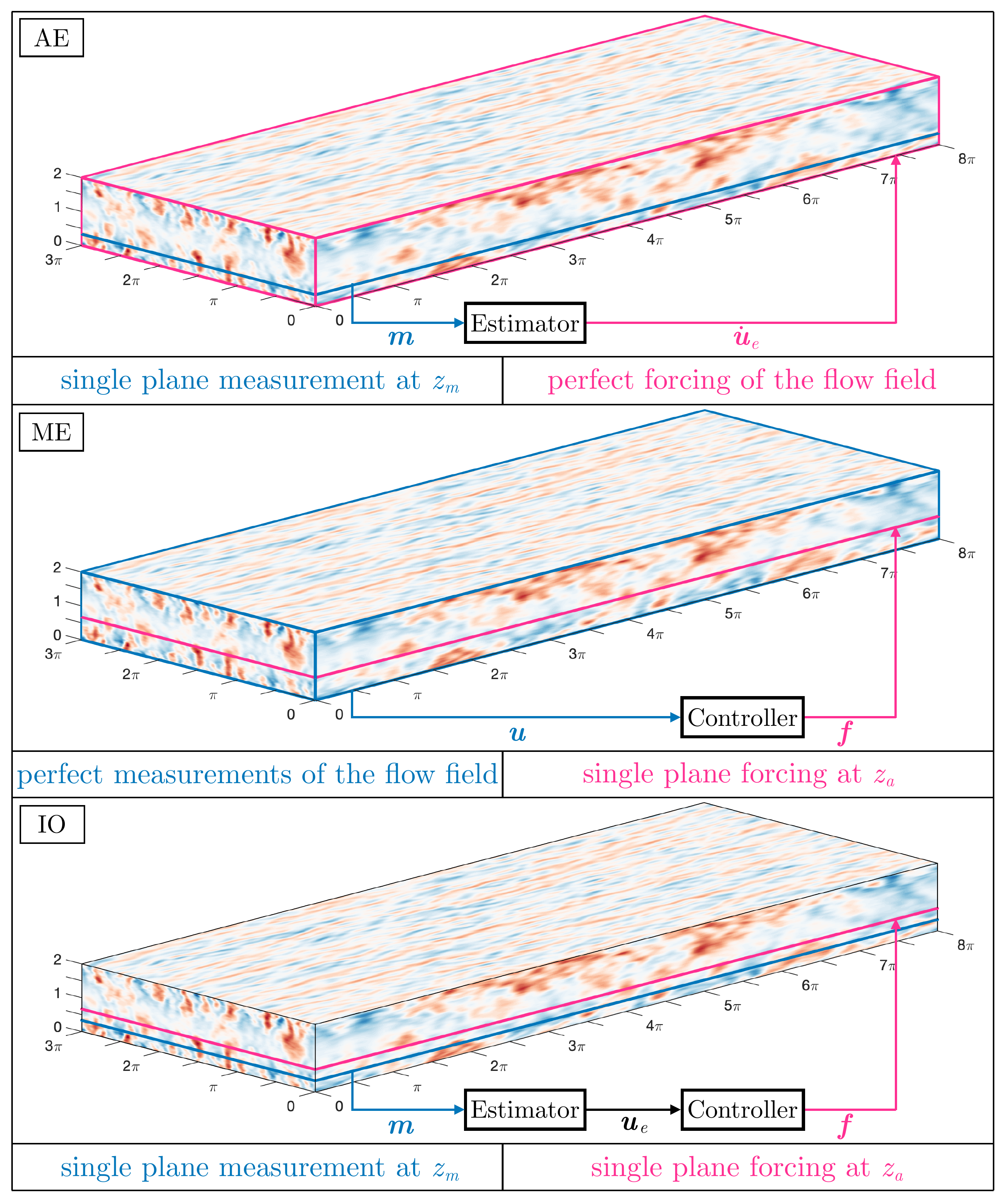}
		\caption[The AE, ME, and IO problems]{The AE, ME, and IO problems.}
		\label{J2:blockOE}\label{J2:block3}
	\end{minipage}
\end{figure}

We now want to use the system $\tilde{ \mathbf P}$, defined in equation \eqref{eqn:Ptilde}, to investigate three different control problems of interest.
The first of these is Actuating Everywhere (AE) control, where the controller can actuate the flow everywhere but is limited to sensors at one wall-normal location. 
The second is Measuring Everywhere (ME) control, where the flow is measured everywhere but now  actuators are limited to one wall-normal location.
The third is Input--Output (IO) control, where sensors and actuators are limited to one wall-normal location. 
This final configuration is of particular interest since it would be the most feasible experimental configuration. 
The three configurations are illustrated in figure \ref{J2:block3}, and the details of their state-space models can be found in \ref{J2:APPDIX:control_ss}.
We study AE, ME and IO because we want to know what price we have to pay when only a single plane of actuators is available (as opposed to actuating the flow everywhere); and what price we have to pay when only a single plane of sensors is available (as opposed to knowledge of the flow everywhere). 
This, in turn, provides insight on the extent to which control of the largest scales is fundamentally difficult; and on the extent to which control is limited by having only a single sensor or a single actuator (per wavenumber pair). 

\subsubsection{Actuating Everywhere (AE) control}

In the Actuating Everywhere (AE) control problem, we can actuate the flow everywhere but only have access to sensor measurements $\hat{\boldsymbol m}$ at a single location $z_s$\footnote{It could also be sensors at various wall heights, actuators at various wall heights, or both.\label{J2:footnote:1}}. These measurements are contaminated by sensor noise $\hat{\boldsymbol n}$. 
The task in the AE problem is to estimate the entire state $\hat{\boldsymbol q}$, and then use the estimate $\hat{\boldsymbol{q}}_e$ to control the flow. \textit{Thus we only have one sensor to measure the flow, and we want to use it to control the flow everywhere.} 

The state estimate is generated using an estimator:
\begin{subequations}
	\begin{align}
	\dot{\hat{\boldsymbol q}}_e(t) &= \left(\mathbf A - \mathbf L \mathbf C_m\right) \hat{\boldsymbol q}_e(t) + \mathbf L\hat{\boldsymbol{m}}(t),
	\\
	\hat{\boldsymbol{u}}_e(t) &= \mathbf C \hat{\boldsymbol{q}}_e(t),
	\label{eqn:LQE}
	\end{align}
\end{subequations}
where $\mathbf L$ is the estimator gain value (designed in \ref{J2:sec:SD}). The estimator knows the dynamics of the system (represented by $\mathbf A $), but it neither knows the initial conditions nor the stochastic disturbances $\hat{\boldsymbol{d}}$ that are applied to the linear operator. It corrects itself using the error between the measurement and its estimate $(\hat{\boldsymbol{m}} - \mathbf C_m {\hat{\boldsymbol q}}_e)$. Finally, the estimated velocity field $\hat{\boldsymbol{u}}_e(t)$ is subtracted from the velocity field itself $\hat{\boldsymbol{u}}(t)$, i.e. the estimate is directly applied as a body force.

\subsubsection{Measuring Everywhere (ME) control}	

In the Measuring Everywhere (ME) control problem, we have an actuator $\mathbf{B}_f
\hat{\boldsymbol f}$ at a single location $z_a$\textsuperscript{\ref{J2:footnote:1}}, and we are given knowledge of the entire system state $\hat{\boldsymbol q}$. 
\textit{Thus we know everything about the flow, but we only have one actuator to control the flow.} 
A controller generates the actuator force $\hat{\boldsymbol f}$:
\begin{align}
\hat{\boldsymbol f}(t) = - \mathbf K \hat{\boldsymbol q}(t), \label{eqn:LQR}
\end{align}
where $\mathbf K$ is the controller gain value (designed in \ref{J2:sec:SD}). 
The `measurement' for this arrangement is the full flow field $\hat{\boldsymbol{q}}$, because it is assumed that the controller `knows everything'.

\subsubsection{Input--Output (IO) control}	
In the Input--Output (IO) control problem, we only have one measurement $\hat{\boldsymbol{m}}$ at $z_s$ available to estimate the flow, and we only have one actuator $\mathbf{B}_f \hat{\boldsymbol{f}}$ at $z_a$ available to control the flow\textsuperscript{\ref{J2:footnote:1}}. The measurement $\hat{\boldsymbol{m}}$, which is contaminated by sensor noise $\hat{\boldsymbol n}$, is used to obtain an estimate  $\hat{\boldsymbol{q}}_e$ (from an estimator),
and the actuator force $\hat{\boldsymbol f}$ is generated with a controller that uses $\hat{\boldsymbol q}_e$. 
\textit{(Thus we only have one sensor to estimate the flow, and we only have one actuator available to control the flow.)} 
To form a combined estimator and controller, we rewrite equation \eqref{eqn:LQR} to include $ \hat{ \boldsymbol q}_e$:
\begin{subequations}
	\begin{align}
	\dot{\hat{\boldsymbol q}}_e(t) &= \left(\mathbf A - \mathbf L \mathbf C_m - \mathbf{B}_f \mathbf K\right) {\hat{\boldsymbol q}}_e(t) + \mathbf L\hat{\boldsymbol{m}}(t),
	\\
	\hat{\boldsymbol f}(t) &= - \mathbf K \hat{\boldsymbol q}_e(t).
	\end{align}
	\label{eqn:LQG}
\end{subequations}

\subsection{Control performance}

We quantify the energy of $\hat{\boldsymbol z}(s)$ with the square of the $\mathcal{H}_2$-norm for one channel half ($0<h\leq 1$)
similar to equation \eqref{J2:Energynorm_u} (\ref{J2:APPDIX:h2norms}). 
 From this, we define $\hat{\mathbf{E}}$, which is the reduction of kinetic energy due to control:

\begin{align}
\hat{\mathbf{E}} = {\frac{ \| \hat{\boldsymbol z}\|_2^2}{ \| \hat{\boldsymbol u}_{ref}\|_2^2}} &= {\frac{ \| \hat{\boldsymbol u}_{ctrl}\|_2^2 +\alpha\|  \hat{\boldsymbol f}\|_2^2}{ \| \hat{\boldsymbol u}_{ref}\|_2^2}} ,
\label{J2:eqn:sigmagammaoe}
\end{align}
where $\|\hat{\boldsymbol u}_{ref}\|_2$ is the $\mathcal{H}_2$-norm of the uncontrolled reference flow, $\|\hat{\boldsymbol u}_{ctrl}\|_2$ the $\mathcal{H}_2$-norm of the controlled flow, and $\|\hat{\boldsymbol f}\|_2$ the $\mathcal{H}_2$-norm of the energy consumed by the actuators. (The parameter $\|\hat{\boldsymbol f}\|_2$ only exists in ME and IO and is treated as negligible in this study.)

\subsection{Optimal sensor and actuator placement}

We want to place the sensors and actuators at the wall height that provides the best performance. To do so, we conduct an iterative minimization search across all possible sensor and actuator locations ($z_s$ and $z_a$) to find the lowest $\hat{\mathbf{E}}$ possible. The iterative gradient minimization employed has been introduced and discussed in earlier studies \citep{Chen2011,Oehler2018}. By following the approach of \citet{oehler2018linearchannell}, it was determined that the optimal collocated placement for the sensor and actuator is at $z_{a} = z_{s} = 0.32$. (Note that only wavenumbers satisfying $|k_x| \leq 0.5$ and $ |k_y| \leq 6$ are considered while computing these optimal placement locations). {$\color{brown}$ The study by \citet{oehler2018linearchannell}} collocates the sensor and actuator (across both channel halves) to simplify the optimisation problem (collocation affects the placement only marginally).

\section{Control performance}
\label{J2:results}
\label{J2:sec:results}

This section is in four parts: \S\ref{J2:sec:across} examines the control performance at individual wavenumber pairs; \S\ref{J2:sec:set} looks at the overall performance; 
\S\ref{J2:sec:wallheights} at the performance across individual wall heights; and \S\ref{J2:sec:forcing} considers the energy consumed by actuation. 

\subsection{Control at individual wavenumber pairs}
\label{J2:sec:across}

\begin{figure}
	\vspace{2mm}
	\centering
	\includegraphics[width=1\textwidth]{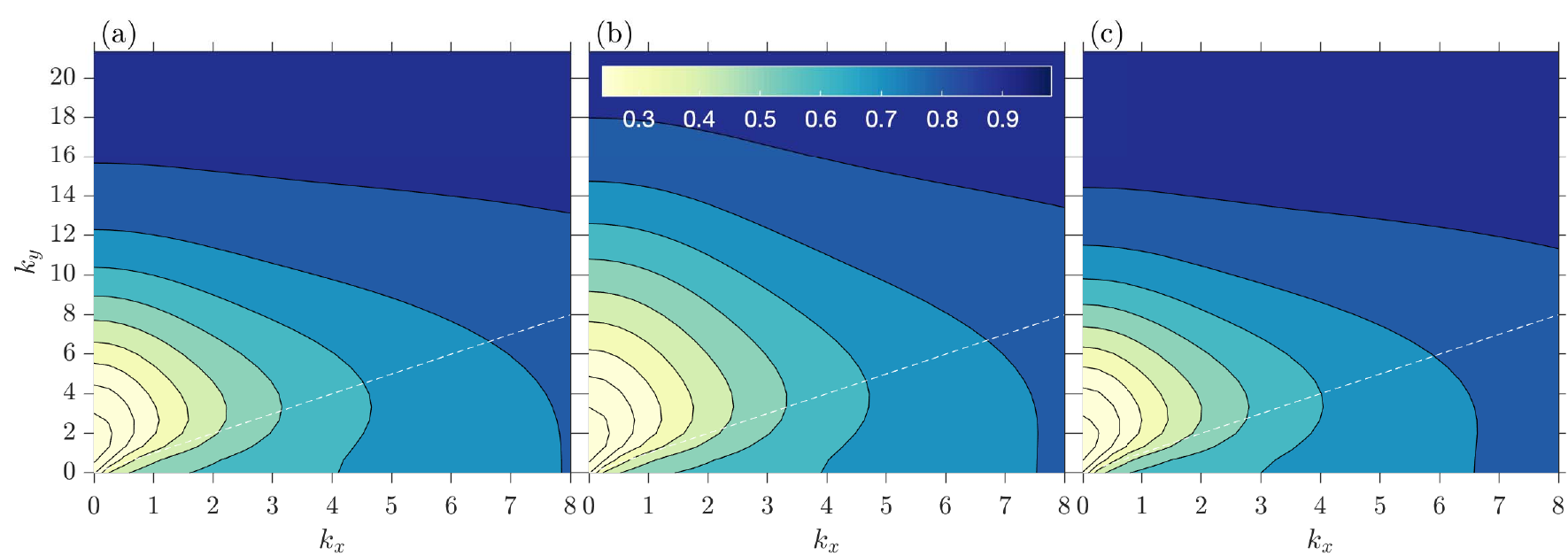}
	\caption[The control-based energy reduction $\hat{\mathbf{E}}_{AE}$, $\hat{\mathbf{E}}_{ME}$ and $\hat{\mathbf{E}}_{IO}$ as a function of $k_x$ and $k_y$.]{The energy reduction $\hat{\mathbf{E}}_{AE}$(a), $\hat{\mathbf{E}}_{ME}$(b) and $\hat{\mathbf{E}}_{IO}$(c) as a function of streamwise $k_x$ and spanwise $k_y$ wavenumber represented by contour levels from $0.24$ (yellow) to $0.98$ (blue).~Also denoted on the figure are the wavenumber pairs where $k_x = k_y(--)$.
		\label{J2:fig:kxkyrange} }
\end{figure}
In this section, we study the control performance of AE, ME and IO over a range of wavenumber pairs ($k_x$,$k_y$). For this purpose, we use the parameter $\hat{\mathbf{E}}$, as defined in equation \eqref{J2:eqn:sigmagammaoe}. In figure \ref{J2:fig:kxkyrange}, $\hat{\mathbf{E}}$ is plotted as a function of $k_x$ and $k_y$ (the channel length is $x = 8 \pi$ and width is $y = 3\pi$; the streamwise resolution is $\Delta k_x = 1/4$ and the spanwise resolution is $\Delta k_y = 2/3$).
The contours of $\hat{\mathbf{E}}$ are almost identical for the three problems. 
Therefore, from figure \ref{J2:fig:kxkyrange}, we see that the performance of the control scenario where we have one optimally placed sensor and actuator (IO) is comparable to the cases where we actuate everywhere (AE) or know everything (ME). 
Hence, we observe that actuating everything does not significantly increase the control performance when we are limited to one sensor. Similarly, measuring everything does not significantly increase the control performance when we are limited to one actuator. 
We observe that, for all three problems, $\hat{\mathbf{E}}$ is the lowest for streamwise-constant structures ($k_x = 0$) with a spanwise wavenumber of $k_y = 4/3$. As the structures become smaller ($k_x$ and $k_y$ increase), $\hat{\mathbf{E}}$ increases. This behavior can partly be explained by the smaller scales being less coherent across wall-normal distances \citep{Madhusudanan2019SLSE}. 		 		
As a consequence, single sensor and actuator control at the smaller scales might not be feasible, even if we consider second-order statistics \citep{zare2017colour} or non-linear controller designs \citep{Lauga2004}.

It is important to assess whether the controllers perform well for the most energetic scales. 	
For this, we compare figure \ref{J2:fig:kxkyrange}, which shows the normalized $\mathcal H_2$-norm for the controlled flow, with figure \ref{J2:fig:kxkyrange_u}, which shows the $\mathcal H_2$-norm of the uncontrolled flow. We observe that, in all three cases, the performance of the controller is the best (low $\hat{\mathbf{E}}$) for the wavenumber pairs ($k_x$,$k_y$) that are most amplified (high $\|\hat{ \boldsymbol u}\|_2^2$). This result is important because it shows that we can reduce the energy of the largest, most amplified scales with a limited number of sensors and actuators.
The same relationship has been observed in similar estimation studies \citep{illingworth2018estimating,oehler2018linear,Madhusudanan2019SLSE}: linear estimation performs best for the wavenumber pairs ($k_x$,$k_y$) that are most amplified (high $\|\hat{ \boldsymbol u}\|_2$).  Therefore, the scales that we can estimate well are also those we can control well.

\subsection{Control in physical space}
\label{J2:sec:results:large}
\label{J2:sec:set}

We now look at control for a set of large-scale structures: $|k_x|\leq 0.5$ and $|k_y|\leq 6$, the range of which is indicated in figure \ref{J2:fig:kxkyrange_u}. The figure shows that these structures are the most amplified in the stochastically forced LNS model, and we can see in figure \ref{J2:fig:kxkyrange} that they are also the best for control.
\label{J2:sec:results_individ}
 
\begin{figure}[]
	\vspace{2mm}
	\centering
	\includegraphics[width=1\textwidth,trim={0cm 4cm 0cm 0cm},clip]{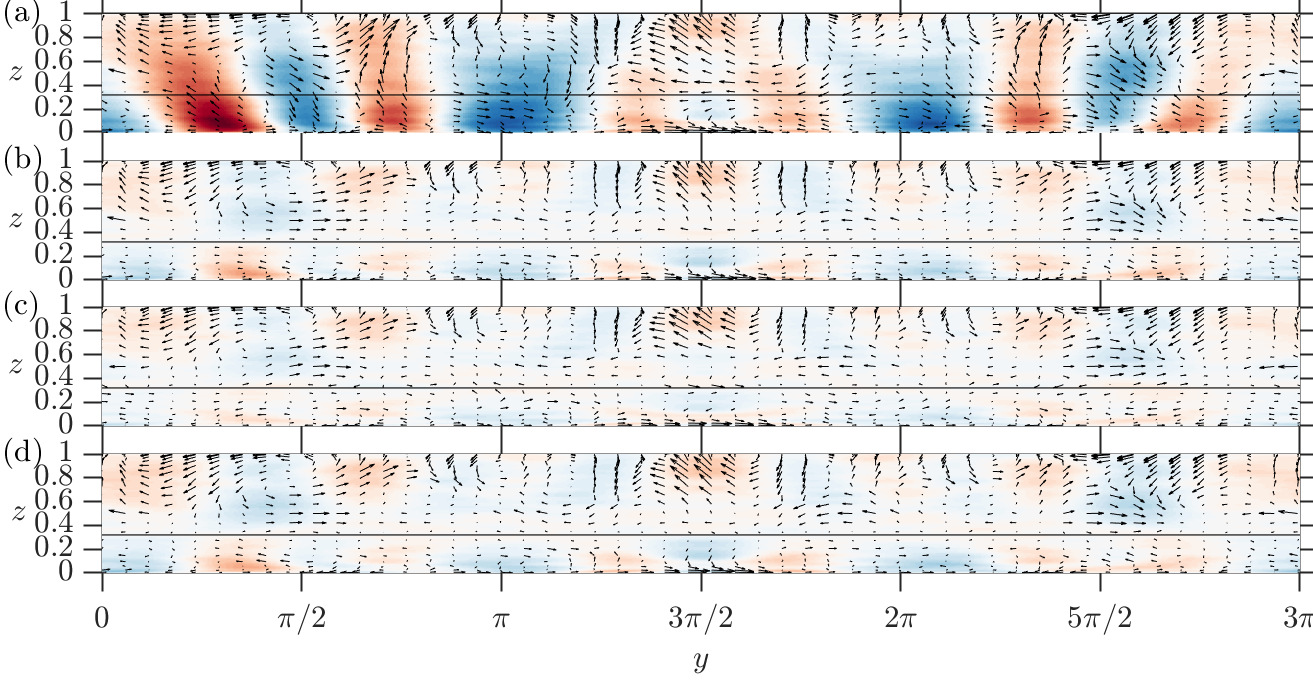}
	\caption[Control of velocity perturbations in physical space]{Velocity perturbations (streamwise: contour; spanwise and wall-normal: vector plot) at $x = 3\pi/2$: (a) uncontrolled reference, (b) AE, (c) ME, and (d) IO. The sensor and actuator are placed at $z_{s} = z_a = 0.32$. An identical scale for the vector plots is employed across all panels. Sixty-five contour levels are shown from $-|u|_{max}$ (blue) to $|u|_{max}$ (red).}
	\label{J2:fig:MEIO}
\end{figure}

We begin by looking at snapshots of the velocity perturbations in two-dimensional planes ($z-y$ at $x = 1.5\pi$) at an instance in time ($t = 0.5$, i.e. after half a channel flow-through). The data is generated from the LM.
Figure \ref{J2:fig:MEIO}a shows the flow field of the uncontrolled (reference) flow. Figures \ref{J2:fig:MEIO}b--\ref{J2:fig:MEIO}d show the controlled flow fields for each of the three cases AE, ME and IO, respectively. 
We observe that all three controllers achieve a significant reduction of the streamwise velocity perturbations everywhere. The spanwise and wall-normal velocity components are also reduced, most notably at $z_s = z_a = 0.32$ (corresponding to the location of the sensors and actuators). 

It is difficult to quantify and compare the control performances from a snapshot in time. 
For that reason, we sum the $\mathcal H_2^2$-norm across all the wavenumber pairs $(|k_x| \leq 0.5,|k_y| \leq 6)$ considered. The parameter $\mathbf{E}_{u,v,w}$ is the ratio of these summed $\mathcal H_2^2$-norms computed from the controlled and the uncontrolled cases, respectively:  
\begin{align}
\mathbf{E}_{u,v,w} &= {\frac{\sum_{i\in k_x, j \in k_y} \| \hat{\boldsymbol z}(i,j)\|_2^2}{\sum_{i\in k_x, j \in k_y} \| \hat{\boldsymbol u}_{ref}(i,j)\|_2^2}}.
\label{J2:eqn:sigmagammaoe2}
\end{align}	
As a consequence, $\mathbf{E}_{u,v,w}$ represents the normalized reduction in kinetic energy due to control integrated across all three velocity components $u$, $v$ and $w$. The values of $\mathbf{E}_{u,v,w}$ are shown in table \ref{J2:tbl:norms}, and they tell us that the overall performance is similar, although ME slightly outperforms AE and IO.
\begin{table}
	\centering
	\begin{tabular}{r|c|c|c}
		%\hline 
	&	${AE}$ & ${ME}$ & ${IO}$ \\
		\hline 
	\hline 
	 $\mathbf{E}_{u,v,w}$ & $0.150$ & $0.135$ & $0.164$
	\\
	\end{tabular} 
	\caption[The control performance for AE, ME and IO]{The control performance for AE, ME and IO. \label{J2:tbl:norms}}
\end{table}

\begin{figure}
	\vspace{2mm}
	\centering
	\includegraphics[width=1\textwidth,trim={3.33cm 12.5cm 0.7cm 3cm},clip]{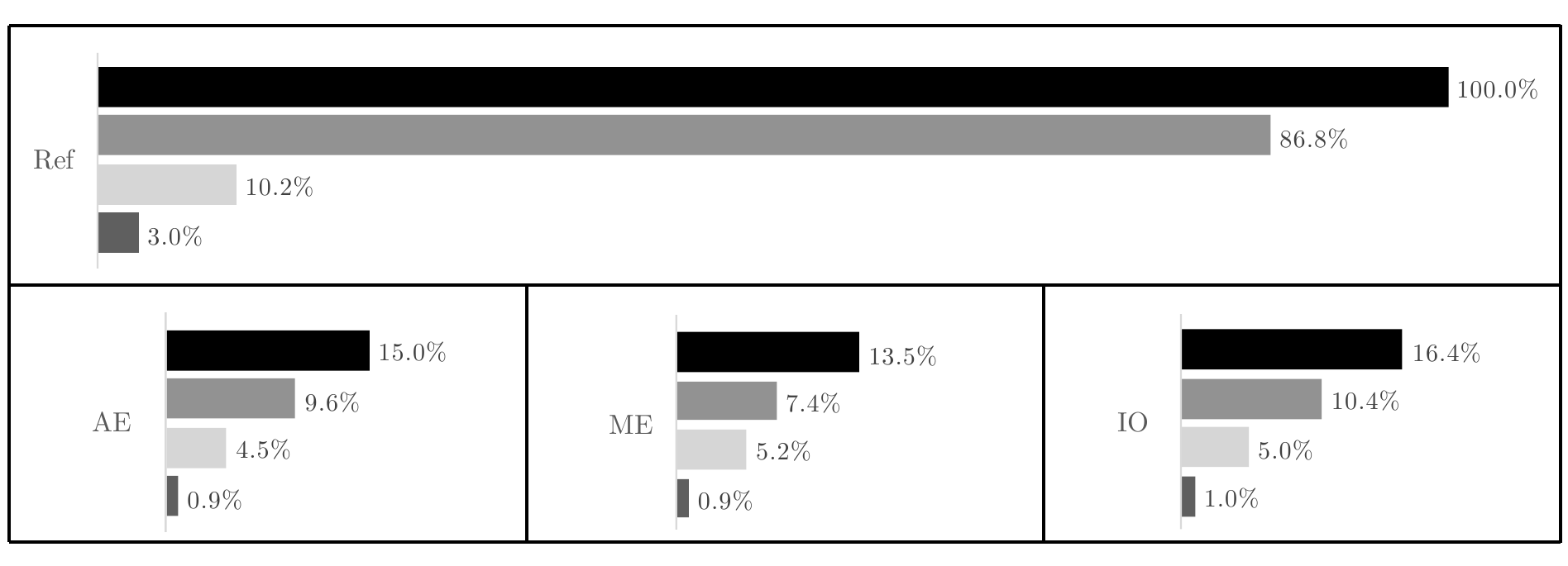}
	\caption[The kinetic energy components for the controlled and uncontrolled flows]{The kinetic energy components for the controlled and uncontrolled flows (relative to the total kinetic energy of the uncontrolled flow): $\mathbf{E}_{u,v,w}$ ({\color{excel1}$\blacksquare$}), $\mathbf{E}_u$({\color{excel2}$\blacksquare$}), $\mathbf{E}_v$({\color{excel3}$\blacksquare$}) and $\mathbf{E}_w$({\color{excel4}$\blacksquare$}), where $\mathbf{E}_{u,v,w} = \mathbf{E}_{u} + \mathbf{E}_{v} +\mathbf{E}_{w}$.}
	\label{J2:excel:components}
\end{figure}

To further understand the control results, it is important to look at the impact of the controllers on each velocity component $[u,v,w]$ separately. Thus, we look at the kinetic energy of each velocity component relative to the energy of the entire uncontrolled flow-field:
\begin{align}
\mathbf{E} = {\frac{\sum_{i\in k_x, j \in k_y} \| \hat{\boldsymbol y}(i,j)\|_2^2}{\sum_{i\in k_x, j \in k_y} \| \hat{\boldsymbol u}_{ref}(i,j)\|_2^2}}.
\end{align}
By setting $\hat{\boldsymbol y}$ to be different velocity components, $\mathbf{E}$ is defined in four different ways: (i) $\mathbf{E}_{u,v,w}$, where $\hat{\boldsymbol y}$ represents all the three velocity components, (ii) $\mathbf{E}_{u}$ where $\hat{\boldsymbol y}$ represents the streamwise velocity component, (iii) $\mathbf{E}_{v}$, where $\hat{\boldsymbol y}$ represents the spanwise velocity component, and (iv) $\mathbf{E}_{w}$, where $\hat{\boldsymbol y}$ represents the wall-normal velocity component.	
Figure \ref{J2:excel:components} shows $\mathbf{E}$ for the uncontrolled reference flow (denoted as Ref) and for the flow subject to AE, ME and IO. In the reference flow, the majority of the energy is contained in $u$ ($87\%$) and the remaining energy in $v$ ($10\%$) and $w$ ($3\%$). After we apply control, we see that, consistent with figures \ref{J2:fig:kxkyrange} and \ref{J2:fig:MEIO} and table \ref{J2:tbl:norms}, the performances of AE, ME and IO are all similar to each other. The overall reduction of energy ($ \mathbf{E}_{u,v,w}$) is $\approx 85\%$, where $ \mathbf{E}_u$ is reduced by $\approx 90\%$, $\mathbf{E}_v$ by $\approx 50\%$ and $ \mathbf{E}_w$ by $\approx 67\%$.  Therefore, the control system is most effective in reducing the streamwise velocity component, which also carries most of the energy.

\subsection{Control across wall heights}
\label{J2:sec:wallheights}

So far, we have looked at the control performance over an entire channel half. It is also important to study the performance of the controllers across wall heights.

\begin{figure}
	\vspace{2mm}
	\centering
	\includegraphics[width=1\textwidth]{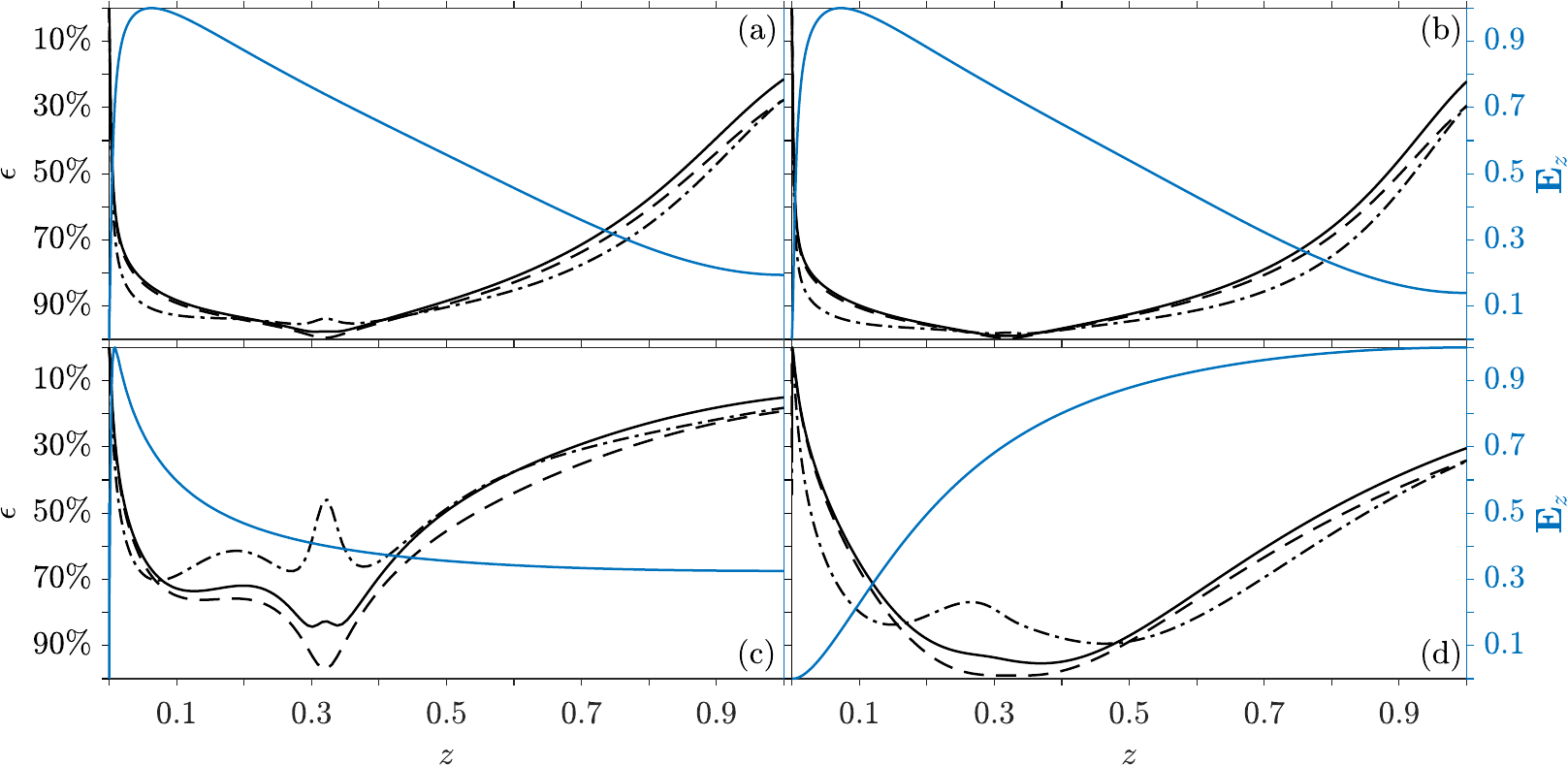}
	\caption[The reduction of kinetic energy as a function of $z$]{Left axis: The reduction of kinetic energy  ($\epsilon_{AE}(--)$,$\epsilon_{ME}(-\cdot-)$ and $\epsilon_{IO}$(---)) as a function of $z$. Right axis: the normalized kinetic energy $\mathbf{E}_z$({\color{blueFig6}---}) as a function of $z$. Results are shown for (a) $[u,v,w]$, (b) $[u]$, (c) $[v]$ and (d) $[w]$.  }
	\label{J2:fig:OEMEIO-RMS}
\end{figure}

For reference, we first compute the normalized kinetic energy of the uncontrolled flow $\mathbf E_z$ as a function of wall-normal location $z$:
\begin{align}
\mathbf E_{z}(z) = \frac{\sum_{i\in k_x, j \in k_y} \|\hat{\boldsymbol y}_{ref}(i,j,z)\|_2^2}{\max (\sum_{i\in k_x, j \in k_y} \|\hat{\boldsymbol y}_{ref}(u,j,z)\|_2^2)}.
\end{align}
Figure \ref{J2:fig:OEMEIO-RMS} shows $\mathbf E_{z}$ as a function of $z$ on the right axis (similar to figure \ref{J2:fig:umaxwallheight}). As in the previous section, the signal $\hat{\boldsymbol{y}}$ represents: all the three velocity components (figure \ref{J2:fig:OEMEIO-RMS}a), the streamwise velocity component (figure \ref{J2:fig:OEMEIO-RMS}b), the spanwise velocity component (figure \ref{J2:fig:OEMEIO-RMS}c), or the wall-normal velocity component (figure \ref{J2:fig:OEMEIO-RMS}d).
From the plot of $\mathbf E_z$ (in blue), we observe that $u$ and $v$ are strongest near the wall (figures \ref{J2:fig:OEMEIO-RMS}b and \ref{J2:fig:OEMEIO-RMS}c), while $w$ is strongest near the channel center (figure \ref{J2:fig:OEMEIO-RMS}d).

We now look at the reduction in the kinetic energy of the controlled flow $\epsilon$ as a function of wall-normal location $z$:
\begin{align}
\epsilon(z) &= 1 - {\frac{\sum_{i\in k_x, j \in k_y} \| \hat{\boldsymbol y}_{ctrl}(i,j,z)\|_2^2}{\sum_{i\in k_x, j \in k_y} \| \hat{\boldsymbol y}_{ref}(i,j,z)\|_2^2}}.
\label{eqn:epsilonnorm2}
\end{align}
There are four different definitions of $\epsilon$ (depending on $\hat{\boldsymbol{y}}$), which are shown in figures \ref{J2:fig:OEMEIO-RMS}a-\ref{J2:fig:OEMEIO-RMS}d on the left axis. 
As before, $\hat{\boldsymbol y}$ represents either all three (figure \ref{J2:fig:OEMEIO-RMS}a) or individual (figure \ref{J2:fig:OEMEIO-RMS}b--d) velocity components.
Parameter $\epsilon$ is shown for AE ($\epsilon_{AE}$), ME ($\epsilon_{ME}$), and IO ($\epsilon_{IO}$). By definition, $\epsilon$ is between $0 \leq \epsilon \leq 1$, where $1$ ($100\%$) indicates the elimination of all kinetic energy and $0$ ($0\%$) indicates that there is no reduction in kinetic energy (for the wavenumber pairs $(|k_x| \leq 0.5,|k_y| \leq 6)$ considered).

From figure \ref{J2:fig:OEMEIO-RMS}a we observe that the performance for all control problems is best near $z_{s} = z_{a} = 0.32$ (where $\epsilon(z)$ is lowest) and decreases with distance from it. A significant reduction of velocity perturbations is observed at all wall heights.  Similar values of $\epsilon$ are achieved in figure \ref{J2:fig:OEMEIO-RMS}b for the streamwise velocity component, which can be explained by $u$ being the most energetic component (figure \ref{J2:excel:components}). AE and IO set $v$ in figure \ref{J2:fig:OEMEIO-RMS}c close to zero around $z_s = z_a = 0.32$. While ME also reduced the energy carried by $v$, the reduction is not as strong as in the case of AE and IO. Additionally, we can see a small influence of the Gaussian-shaped actuator on the results in ME and IO. Figure \ref{J2:fig:OEMEIO-RMS}d shows that all three problems set the wall-normal velocity close to zero at one wall height. The transport of momentum in the vicinity of this wall height is attenuated, which prevents the formation of streamwise structures \citep{toh2005interaction}. This mechanism is employed in opposition-controlled wall-bounded flows \citep{hammond1998observed,luhar2014opposition,nakashima2017assessment}, where the controller is specifically designed to create a plane of zero wall-normal momentum that is referred to as a ``virtual wall''. We did not choose an opposition control design but instead selected a general cost function to reduce velocity perturbations everywhere. Since the three $\mathcal{H}_2$-optimal control designs seem to all create a ``virtual wall'', the results suggest that this approach is the most effective one in the control of turbulent channel flows utilizing single-plane sensors and single-plane actuators. 

Let us compare $\epsilon_{ME}$, where the flow field is known everywhere, to $\epsilon_{IO}$, where only one location is known. We see that ME performs marginally better than IO everywhere outside the vicinity of the sensor at $z = 0.32$.  This suggests that IO is focusing its control efforts on the region near $z = 0.32$ (that it `knows well') at the expense of a slight reduction in control performance everywhere else.

If we compare $\epsilon_{AE}$, where actuation is provided everywhere, to $\epsilon_{IO}$, where actuation is provided at only one location, we can see that they are almost identical to each other except in the vicinity of the single actuator at $z = 0.32$. Therefore, near $z = 0.32$, the performance of IO  must be primarily limited by the single actuator; while at all other locations its performance is limited by the single sensor.	

Finally, by comparing $\epsilon_{AE}$ with $\epsilon_{ME}$, we can conclude that feedback control overall must be slightly more limited by the single sensor than the single actuator.

\subsection{Control forces}
\label{J2:sec:forcing}

So far, we have studied the effect that the three control problems have on the velocity perturbations. Each problem continuously forces the flow to prevent perturbations from growing. In this section, we study these continuous forces.		
In particular, we look at the percentage of the forcing that is applied to the streamwise, spanwise and wall-normal directions.
One may ask how it is possible to look at the distribution of actuation forces, despite having almost no actuation cost (i.e. $\alpha$ is relatively small and $\boldsymbol{f}$ is relatively large). The answer lies with the cost function (\ref{J2:APPDIX:cost}), which will still prioritise actuation in the flow direction that gives the best results for the least amount of energy.

In figure \ref{J2:excel:forcing}, we plot the energy consumed
by $f_x$, $ f_y$ and $ f_z$ as a percentage of the total $\boldsymbol{f}$, which we refer to as $\mathbf{E}_{ f_x}$, $\mathbf{E}_{ f_y}$ and $\mathbf{E}_{ f_z}$ (see  \ref{J2:APPDIX:h2norms} for the $\mathcal H_2$-norms):	
\begin{align}
\mathbf{E}_f = {\frac{\sum_{i\in k_x, j \in k_y} \|  \hat{f}(i,j)\|_2^2}{\sum_{i\in k_x, j \in k_y} \| \hat{\boldsymbol f}(i,j)\|_2^2}}.
\label{J2:eqn:Ef}
\end{align}
We observe that in AE, which actuates the flow everywhere, the largest forcing component is $\mathbf{E}_{ f_x}$ (streamwise), and the smallest forcing component is $\mathbf{E}_{ f_z}$ (wall-normal).
In ME and IO, which actuate the flow at only one location, the largest forcing component is $\mathbf{E}_{ f_z}$ (wall-normal) and the smallest forcing component is $\mathbf{E}_{ f_x}$ (streamwise). 

We can explain the observed result using two mechanisms:

\begin{figure}
	\vspace{2mm}
	\centering
	\includegraphics[width=1\textwidth,trim={0.01cm 0.01cm 0.01cm 0.01cm},clip]{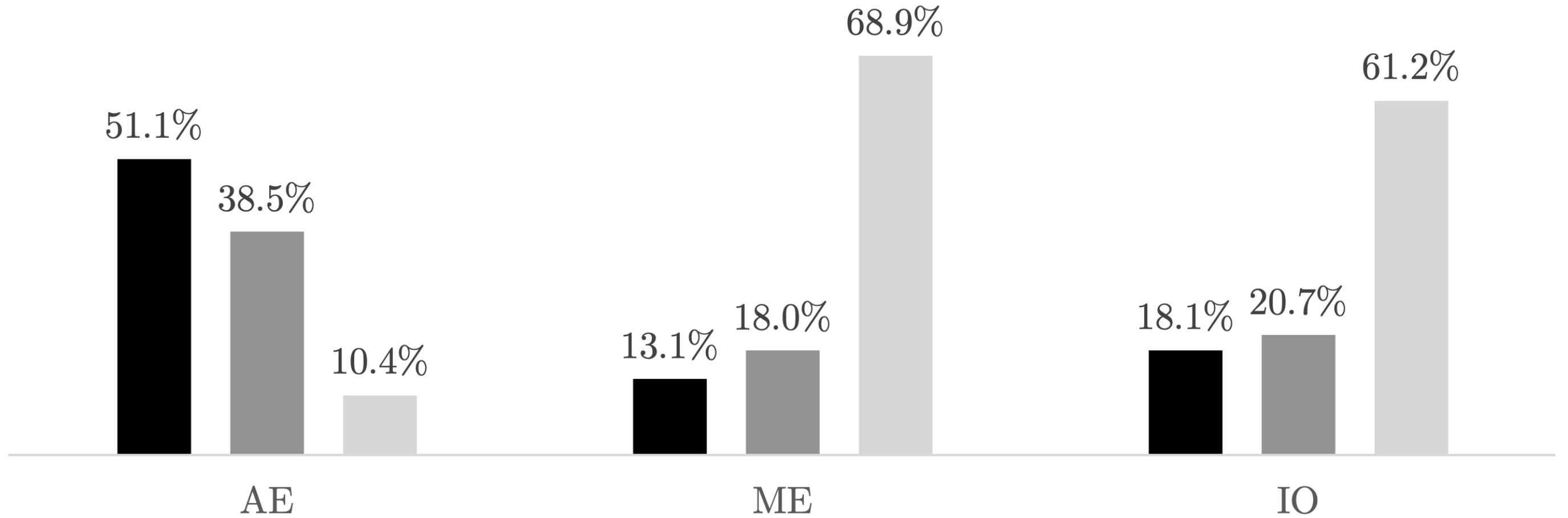}
	\caption[The distribution of forcing]{The distribution of forcing between $\mathbf{E}_{ f_x}$ ({\color{excel1}$\blacksquare$}), $\mathbf{E}_{ f_y}$({\color{excel2}$\blacksquare$}) and $\mathbf{E}_{ f_z}$({\color{excel3}$\blacksquare$}), where $\mathbf{E}_{ f_x} + \mathbf{E}_{ f_y} + \mathbf{E}_{ f_z} = 1$.}
	\label{J2:excel:forcing}
\end{figure}
\begin{itemize}
	\item[(i)] Direct elimination: velocity perturbations are counter-perturbed as soon as they are detected, which 
	is mostly employed by AE. One may ask why AE only allocates $\mathbf{E}_{f_x}  \approx51\%$ of energy to $f_x$ even though the energy reduction in the streamwise direction is responsible for $\approx91\%$ of the overall energy reduction. The answer is that, once we apply control, streamwise perturbations are not given a chance to amplify, which allows the controller to allocate more energy to $f_y$ and $f_z$.
	\item[(ii)] Indirect elimination: is used for wall heights at which actuation is not available.  As soon as velocity perturbations are detected, the actuator introduces counter-perturbations in the wall-normal direction. These counter-perturbations help to suppress the streamwise vorticity perturbations that give rise to the energetic streamwise velocity perturbations. The indirect elimination technique is employed by ME and IO, and explains their high allocation of energy to $ f_z$ ($\mathbf{E}_{f_z} = 68.9\%$ in ME and $\mathbf{E}_{f_z} = 61.2\%$ in IO). The streamwise $f_x$ and spanwise $f_y$ forces primarily affect control locally around the actuator location and as a consequence are given less priority. 
	
	Indirect elimination also explains the peaks observed in figure \ref{J2:fig:OEMEIO-RMS} around the sensor and actuator location $(z=0.32)$ for IO and particularly for ME.	
\end{itemize}

\subsection{Individual control directions}
\label{J2:sec:IndvControl}

\begin{figure}
	\vspace{2mm}
	\centering
	\includegraphics[width=1\textwidth,trim={3.33cm  7.5cm 0.7cm 3cm},clip]{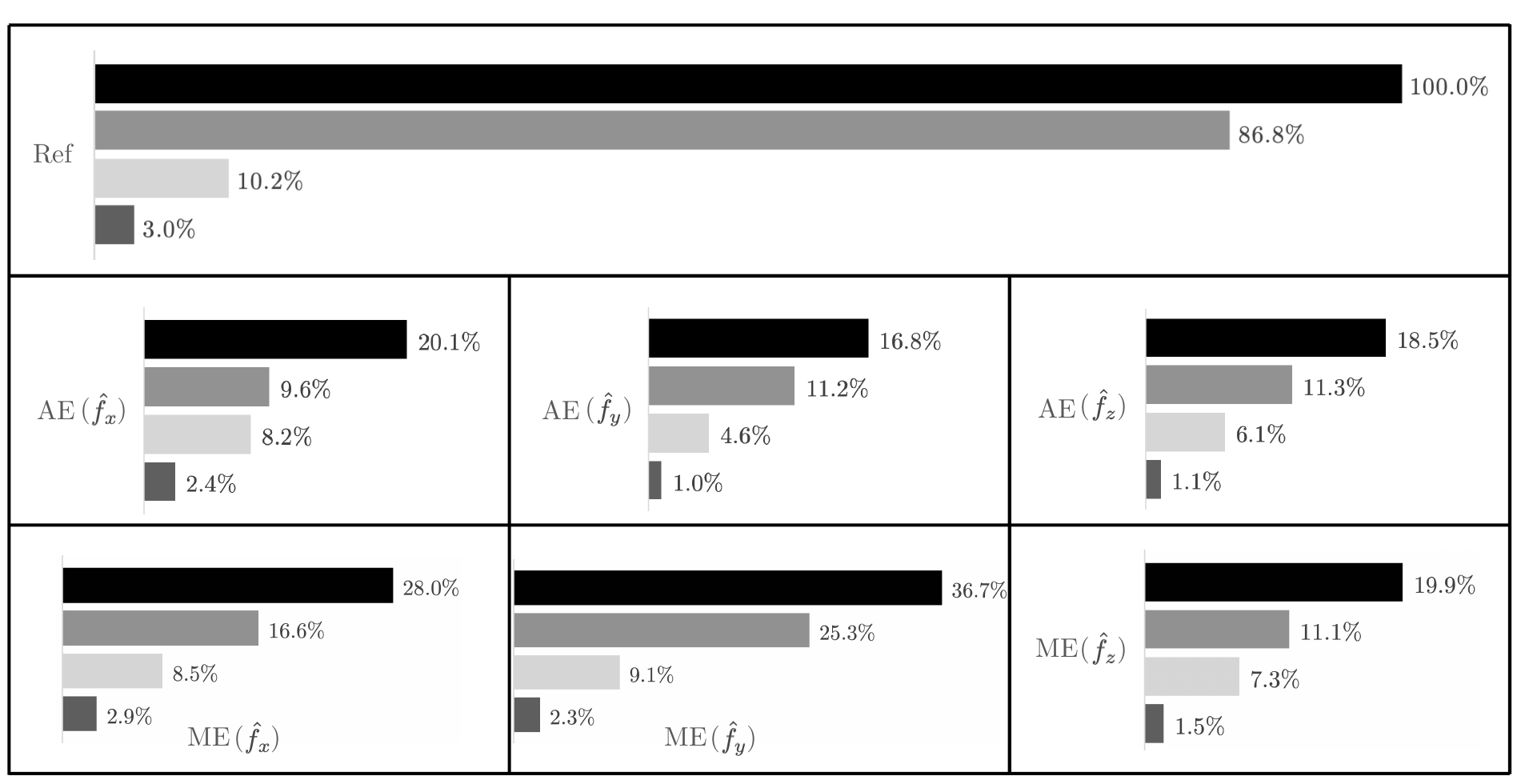}
	\caption[The kinetic energy components for the controlled and uncontrolled flows (relative to the total kinetic energy of the uncontrolled flow) for single direction AE and ME]{The kinetic energy components for the controlled and uncontrolled flows (relative to the total kinetic energy of the uncontrolled flow): $\mathbf{E}_{u,v,w}$ ({\color{excel1}$\blacksquare$}), $\mathbf{E}_u$({\color{excel2}$\blacksquare$}), $\mathbf{E}_v$({\color{excel3}$\blacksquare$}) and $\mathbf{E}_w$({\color{excel4}$\blacksquare$}), where $\mathbf{E}_{u,v,w} = \mathbf{E}_{u} + \mathbf{E}_{v} +\mathbf{E}_{w}$. Results are shown for (AE), where we either force the streamwise direction ($\hat{f}_x$) everywhere, the spanwise direction ($\hat{f}_y$) everywhere, or wall-normal direction ($\hat{f}_z$) everywhere; and results are shown for (ME), where we limit the plane of actuators to either force the streamwise ($\hat{f}_x$), spanwise ($\hat{f}_y$) or wall-normal direction ($\hat{f}_z$).}
	\label{J2:excel:components2}
\end{figure}

In the previous section, we looked at the distribution of the control forces in the three flow directions. This was possible because the cost function prioritises the forcing component for which actuation is most effective. We now look at each forcing component independently, to see their individual effectiveness. Therefore, we repeat the results of figure \ref{J2:excel:components} for AE and ME in figure \ref{J2:excel:components2}. For Actuation Everywhere (AE) control, we force the flow everywhere in either the streamwise ($\hat{f}_x$), spanwise ($\hat{f}_y$) or wall-normal flow direction ($\hat{f}_z$). For Measurements Everywhere (ME) control, we limit the actuator to either the streamwise ($\hat{f}_x$), spanwise ($\hat{f}_y$) or wall-normal direction ($\hat{f}_z$). We do not show the results for IO, because they are very similar to ME.

Forcing in the streamwise direction everywhere AE($\hat{f}_x$) achieves the best reduction in energy in the streamwise flow direction. In fact, it is indistinguishable from the results in figure \ref{J2:excel:components}. However, AE ($\hat {f}_x$) does not significantly affect the spanwise or wall-normal velocity fluctuations. The reduction of streamwise energy for the remaining two cases AE($\hat{f}_y$) and AE($\hat{f}_z$) is not as high as for AE($\hat{f}_x$) but they achieve a better reduction of energy in the spanwise and wall-normal flow directions. Overall, the three cases perform similarly, with AE($\hat{f}_y$) slightly outperforming AE($\hat{f}_z$) and AE($\hat{f}_x)$.

Actuating in the streamwise direction at one wall height while measuring the flow everywhere ME ($\hat{f}_x$), performs better than just forcing the spanwise direction ME ($\hat{f}_y$) but worse than just forcing the wall-normal direction ME ($\hat{f}_z$). The performance differences between the flow directions in ME are greater than for AE, where ME($\hat{f}_z$) outperforms ME($\hat{f}_x$) and ME($\hat{f}_y$) significantly. Consequently, the results in figure \ref{J2:excel:components2} highlight the importance of wall-normal velocity fluctuations for effective control of the very large-scale structures considered. Control in the wall-normal direction has been utilised in many previous studies where it has shown to be effective over a range of scales (e.g.~\cite{choi1994active,lee1998suboptimal,lim2004singular,sharma2011relaminarisation,luhar2014opposition,toedtli2019predicting}). The interaction between wall-normal velocity fluctuations and the mean shear is responsible for energy extraction from the mean flow and can energize streamwise velocity fluctuations. It also explains why $z = 0.32$ is the optimal sensor location rather than the location of the peak in energy ($z \approx 0.06$). At $z = 0.32$, the actuators can reach energetic regions of the flow near the wall, while also being able to influence the remaining less energetic regions (see kinetic energy $\mathbf{E}_z$ in figure \ref{J2:fig:OEMEIO-RMS}).

Despite ME($\hat{f}_z$) outperforming ME($\hat{f}_x$) and ME($\hat{f}_y$), we can still see a notable reduction in energy (of $72$\% for $\hat{f}_x$ and of $63$\% for $\hat{f}_y$). This is due to the creation of a ``virtual wall'', as explained in \S \ref{J2:sec:wallheights}.
The linear model that we form about the mean velocity profile, despite being stable, exhibits transient growth (of up to an order of magnitude \citep{DelAlamo2006,Pujals2009}). Through transient growth, small random disturbances can grow into significant velocity fluctuations. Optimal transient growth gives rise to streamwise velocity steaks that are created by initialising the linear model with counter-rotating streamwise vortices filling the entire channel height \cite{Pujals2009}. These amplification mechanisms are interrupted through the establishment of this  ``virtual wall''.

\section{Conclusions}
\label{J2:sec:discussion}

We have considered linear feedback control of a turbulent channel at Re$_\tau = 2000$ using the linearized Navier-Stokes equations (LNS) which are formed about the turbulent mean. The linear operator is augmented with an eddy viscosity (following many previous studies) and is assumed to be stochastically forced.
Applying any type of control will alter the mean velocity profile and with it the linear model itself. As a consequence, any controlled states cannot be fully described with the present approach. However, employing the LNS equations still provides insight into control, without the requirement of running costly DNS or experimental studies.

The particular focus was on three control problems: (i) AE, where measurements are limited to one optimal wall-normal location, but actuation is available everywhere; (ii) ME, where actuators are limited to one optimal wall-normal location, but measurements are available everywhere; and (iii) IO, where sensors and actuators are limited to one optimal wall-normal location. All three problems performed similarly. From these results we can infer that measuring everywhere does not significantly increase the control performance when we are limited to one actuator location. Likewise, actuating everywhere does not significantly increase the control performance when we are limited to one sensor location. 
Our three control problems perform best for the largest scales that (i) are high in energy when stochastically forced, (ii) exhibit large transient growth and (iii) are coherent over large wall-normal distances. Therefore, we choose to look at a specific range of wavenumbers ($|k_x|\leq 0.5$ and $|k_y|\leq 6$), corresponding to the largest scales, in more detail.
We saw an overall reduction in kinetic energy of $\approx85\%$, where the streamwise velocity component was most attenuated (by $\approx90\%$). 
To further analyze the largest scales, we looked at the effect of control at individual wall heights. The performance was best near the sensor and actuator location ($z = 0.32$, which was based on the optimal placement results of \citet{oehler2018linearchannell}) and deteriorated with distance from it. 
The final part studied the distribution of the forcing between the streamwise $f_x$, spanwise $f_y$ and wall-normal $f_z$ components. For AE, $f_x$ was strongest and $f_z$ weakest, while for ME and IO, $f_z$ was strongest and $f_x$ weakest. AE, which forces the flow everywhere, relies on directly eliminating structures as soon as they are detected, which is why it prioritizes streamwise forcing $f_x$. Meanwhile ME and IO, which only force the flow at a single location, mainly employ wall-normal forcing ($f_z$), thereby eliminating velocity perturbations by leveraging the mean wall-normal shear and establishing a virtual wall.

\section*{Acknowledgements}
The authors would like to thank Sean Symon and Anagha Madhusudanan for their extensive feedback and support during the creation of this paper and are grateful for the financial support of the Australian Research Council.

\appendix

\section{Spectral discretisation of the channel equation}
\label{J2:apdx:1}

We generate the eddy viscosity profile and mean velocity profile (equation \eqref{J2:eqn:eddyprofile}) for one channel half using Chebyshev collocation of order $N_\nu = 200$ \citep{trefethen2000spectral}. Barycentric Lagrange interpolation \citep{Berrut2004bary} is used to map the results to both channel halves.
For the main channel flow (equation \eqref{J2:eqn:ssp}) we employ Chebyshev collocation of order $N_c = 200$. When looking at results for one channel half, we employ barycentric interpolation to map the outputs onto a Chebyshev grid of order $N_{out} = 200$. 
We apply stochastic forcing, which is white in wavenumber space and time, at each grid point $i$ with a covariance $\mathbb{E}( \hat{d}_i \hat{{d}_i^*}) = 1$, where $\mathbb{E}$ the expected value.

\subsection{Convergence}
\label{J2:sec:converg}

To check for convergence of results in $N_c$, we look at $\mathbf{E}_{u,v,w}$ (see table \ref{J2:tbl:norms}). Increasing $N_c$ from $200$ to $400$ changes the result by $0.45$\% for IOC, $0.5$\% for AE and $0.57$\% for ME. If we look at the energy of the uncontrolled flow ($\sum_{i\in k_x, j \in k_y} \| \hat{\boldsymbol u}_{ref}(i,j)\|_2^2$), we observe a change of $0.11$\%.

\section{Control}
\label{J2:sec:CO}
\label{J2:sec:SD}
\label{J2:APPDIX:control_ss}
\label{J2:APPDIX:h2norms}
\subsection{Control objective}
\label{J2:APPDIX:cost}
The following cost function defines the control objective $\hat{\boldsymbol z}$ (equation (\ref{eqn:Ptilde}b)) and is used for the $\mathcal H_2$-optimal control problems:
\begin{align}
J = \mathbb{E} \left\{\lim\limits_{t\rightarrow\infty} \frac{1}{T}\int_{0}^{T}\left(  \int_{0}^{h} \hat{\boldsymbol u}(z,t)^*\hat{\boldsymbol u}(z,t) dz + {\alpha^2} \hat{\boldsymbol f}(t)^*\hat{\boldsymbol f}(t) \right) dt\right\},
\label{eqn:CostFunc1}
\end{align} 
where

\begin{align}
\int_{0}^{h} \hat{\boldsymbol u}(z,t)^*\hat{\boldsymbol u}(z,t) dz + {\alpha^2} \hat{\boldsymbol f}(t)^*\hat{\boldsymbol f}(t)
&\equiv  \begin{bmatrix}\mathbf M^{1/2} \mathbf C \hat{\boldsymbol q}(t) & \mathbf{0} \\ \mathbf{0} & \alpha \hat{\boldsymbol f}(t)\end{bmatrix}^*\begin{bmatrix}\mathbf M^{1/2} \mathbf C \hat{\boldsymbol q}(t) & \mathbf{0} \\ \mathbf{0} & \alpha \hat{\boldsymbol f}(t)\end{bmatrix}  \nonumber \\&\equiv  \left[ \hat{\boldsymbol z}(t)\right]^*\left[ \hat{\boldsymbol z}(t)\right]. 
\end{align}

	\subsection{The cost of actuation}
	
\label{J2:sec:noiseenergy}
	
\begin{figure}
	\vspace{2mm}
	\centering
	\includegraphics[width=1\textwidth,clip]{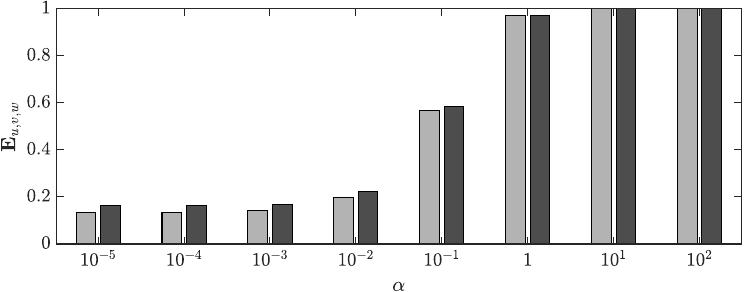}
	\caption[]{The overall control performance ($\mathbf{E}_{u,v,w}$) as a function of control cost ($\alpha$) for ME({\color[rgb]{0.7,0.7,0.7}$\blacksquare$}) and IO({\color[rgb]{0.3,0.3,0.3}$\blacksquare$}).}
	\label{J2:NoiseStudy}
\end{figure}

	To determine the best control performance possible, control needs to be insensitive to $\alpha$. We achieve this by setting $\alpha = 10^{-4}$. To show that control is in fact insensitive to this choice $\alpha$, we plot the control performance for a set of $\alpha$ in Figure \ref{J2:NoiseStudy} when either ME control or IO control is active. 
		The results show that if $\alpha < 10^{-2}$ control is insensitive to $\alpha$ and provides the best result possible. If $\alpha > 10^{0}$, control becomes too expensive in the cost function (equation \ref{eqn:CostFunc1}) and the controller decides to do nothing, which results in $\mathbf{E}_{u,v,w} \approx 1$.

\subsection{The estimator and controller gain matrices}

The gain matrix $\mathbf L$ for AE is designed by solving the following Riccati equation for $\mathbf Y$:
\begin{subequations}
	\begin{align}
	\mathbf{AY} + \mathbf{YA}^* - \mathbf{YC}_m^* \mathbf{V}^{-1} \mathbf{C}_m \mathbf{Y} + \mathbf B_d \mathbf B_d^* = 0,
	\\
	\mathbf L  = \mathbf Y \mathbf C_m^* \mathbf V^{-1}.
	\end{align}
\end{subequations}

The gain matrix $\mathbf K$ for ME is designed by solving the following Riccati equation for $\mathbf X$:
\begin{subequations}
	\begin{align}
	\mathbf A^* \mathbf X + \mathbf{XA} - \mathbf X \mathbf B_f ( \alpha^2\mathbf{I})^{-1} \mathbf B_f^* \mathbf X + \mathbf C_z^* \mathbf C_z = 0,
	\\
	\mathbf K  = ( \alpha^2\mathbf{I})^{-1} \mathbf B_f^* \mathbf X,
	\end{align}
\end{subequations}
where $\mathbf{I}$ is the identity matrix.
The principle of separation for estimation and control states that the independently designed $\mathbf L$ and $\mathbf K$ are still optimal when combined \citep{kalman1960contributions}. Therefore, we do not have to find them again for IO.

\subsection{State-space model}

\begin{figure}
	\vspace{0 mm}
	\centering
	\includegraphics[width=0.75\linewidth]{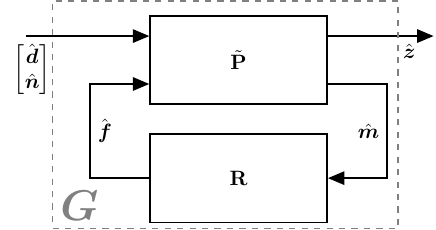}	
	\caption[Block diagram of $\mathbf G$]{Block diagram of $\mathbf G$.\label{fig:blockgen3}}
\end{figure}
\begin{figure}
	\centering
	\includegraphics[width=0.6\textwidth,trim={13cm 35cm 24cm 13.5cm},clip]{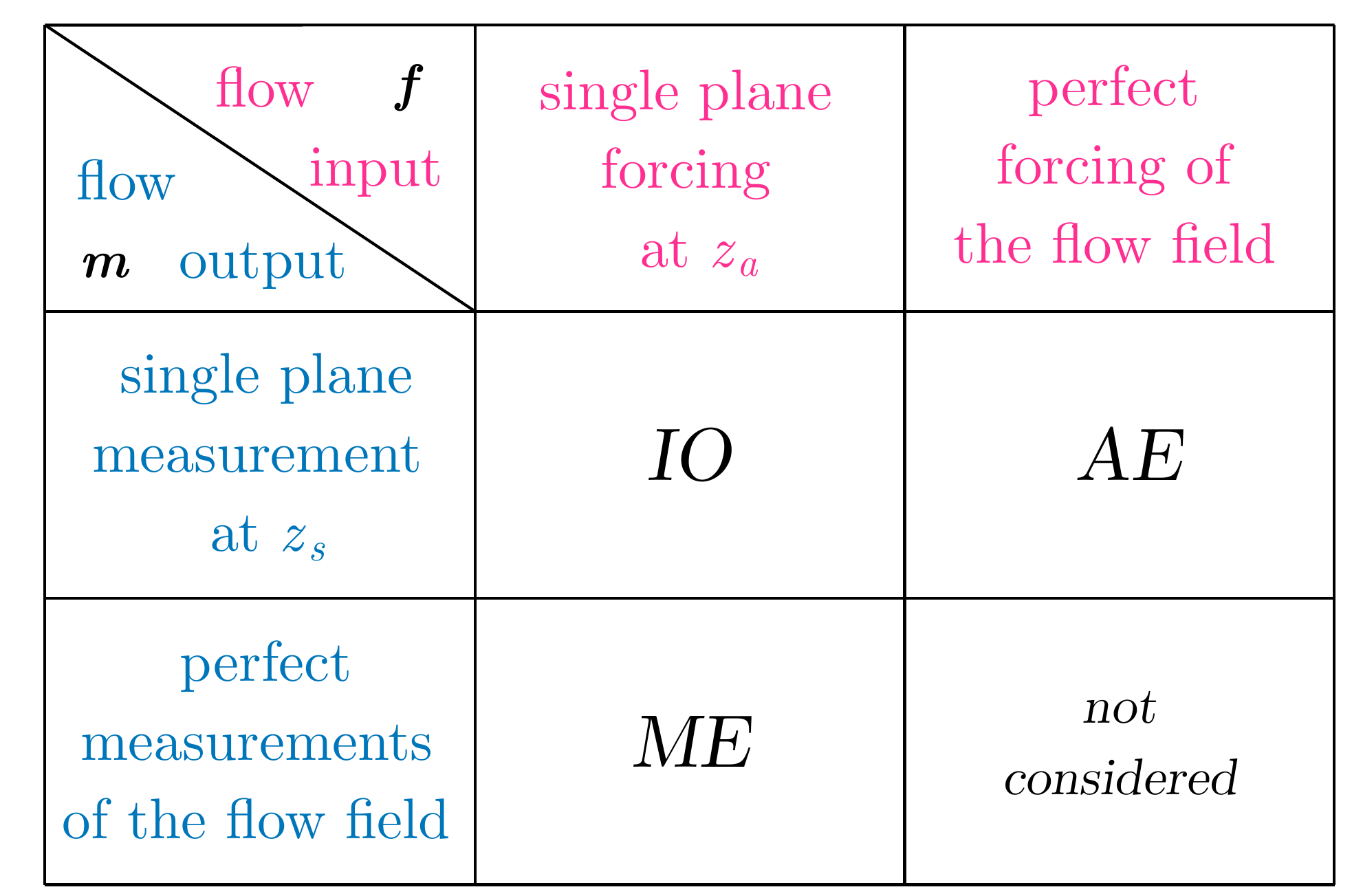}
	\caption[Inputs and outputs of the control problems]{Inputs and outputs of the control problems.\label{J2:block4b}}	
\end{figure}

The AE, ME and IO problems introduce a secondary system  $\mathbf R$ to the flow $\tilde{\mathbf P}$ (figure \ref{fig:blockgen3}), where $\mathbf R$ is either an estimator, a controller or both (figure \ref{J2:block3}). To quantify the control performance of the three problems, we need to express the feedback interconnection of $\tilde{\mathbf P}$ and $\mathbf R$ as a single transfer function.

The measurement signal $\hat{\boldsymbol m}$ acts as an input and the force signal $\hat{\boldsymbol{f}}$ as an output to the secondary system:
\begin{align}
\hat{\boldsymbol f}(t) =  \mathbf R (t) \hat{\boldsymbol{m}}(t). \label{J2:secondary}
\end{align}
The signals $\hat{\boldsymbol{m}}$ and $\hat{\boldsymbol{f}}$ depend on the problem we consider (figure \ref{J2:block4b}). By substituting $\mathbf R\hat{\boldsymbol{m}}$ for $\hat{\boldsymbol{f}}$ in $\tilde{ \mathbf P}$ (equation \eqref{eqn:Ptilde}), we can form the overall state-space model $\mathbf {G}$ (figure \ref{fig:blockgen3}), using a linear fractional transformation (LFT) \citep{astrom2010feedback}:

\begin{minipage}{.55\textwidth}
	\vspace{0 mm}
	\centering
	\includegraphics[width=1\linewidth]{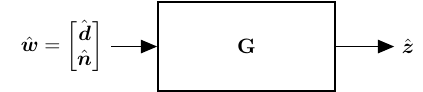}
	\label{fig:blockG}
\end{minipage}	
\begin{minipage}{.4\textwidth}
	\vspace{-12 mm}
	\begin{subequations}
		\begin{align}
		\dot{\hat{\boldsymbol x}} &= \mathbf{A}_L \hat{\boldsymbol{x}} + \mathbf{B}_L \hat{\boldsymbol{w}},
		\\
		\hat{\boldsymbol{z}} &=  \mathbf{C}_L \hat{\boldsymbol{x}},
		\end{align}
	\end{subequations} 
\end{minipage}
where $\mathbf A_L \hat{\boldsymbol{x}}$ describes the state dynamics, $\mathbf B_L \hat{\boldsymbol{w}}$ the input dynamics and $\mathbf C_L\hat{\boldsymbol{x}}$ the output dynamics of the LFT.

To form the LFT for AE we ignore $\hat{\boldsymbol f}$ in $\tilde{ \mathbf P}$ (equation \eqref{eqn:Ptilde}) and directly apply $\dot{{\boldsymbol q}}_e$ (equation \eqref{eqn:LQE}) to $\dot{{\boldsymbol q}}$. The state-space model of $\mathbf{G}_{AE}(t)$ is:
\begin{align}
\dot{\hat{\boldsymbol{q}}} &= \left(\mathbf{A} -  \mathbf L \mathbf C_m  \right) \hat{\boldsymbol{q}} + \begin{bmatrix}\mathbf B_d &  - \mathbf L\mathbf V^{1/2}  \end{bmatrix}   \begin{bmatrix} \hat{\boldsymbol{d}} \\ \hat{\boldsymbol n} \end{bmatrix}, 
\\
\hat{\boldsymbol z} &= \mathbf C_z  \hat{\boldsymbol{q}}.
\end{align}

To form the LFT for ME we ignore $\hat{\boldsymbol m}$ in $\mathbf{\tilde P}$ (equation \eqref{eqn:Ptilde}) and directly form $\hat{\mathbf f}$ from $\hat{\boldsymbol{q}}$ (equation \eqref{eqn:LQR}). The state-space model of $\mathbf{G}_{ME}$ is:
\begin{align*}
\dot{\hat{\boldsymbol q}}
&=
\left(
\mathbf A - \mathbf B_f \mathbf K 
\right)
\hat{\boldsymbol q}
+
\mathbf B_d
\hat{\boldsymbol d},
\\
\hat{\boldsymbol z}
&=
\begin{bmatrix}
\mathbf C_z
\\
- \alpha\mathbf K
\end{bmatrix}
\hat{\boldsymbol q}.
\end{align*}

To form the LFT for IO we combine $\mathbf R$ (equation \eqref{eqn:LQG}) with $\mathbf{\tilde P}$ (equation \ref{eqn:Ptilde}). The state-space model of $\mathbf{G}_{IO}$ is:

\begin{align*}
\begin{bmatrix}
\dot{\hat{\boldsymbol q}} \\ \dot{\hat{\boldsymbol q}}_e
\end{bmatrix}
&=
\begin{bmatrix}
\mathbf A & -\mathbf B_f \mathbf F  \\
\mathbf L \mathbf C_m & \mathbf A - \mathbf B_f \mathbf F - \mathbf L \mathbf C_m  \\
\end{bmatrix}
\begin{bmatrix}
\hat{\boldsymbol q} \\ \hat{\boldsymbol q}_e
\end{bmatrix}
+
\begin{bmatrix}
\mathbf B_d & \mathbf 0 \\
\mathbf 0 & \mathbf L \mathbf V^{1/2} 
\end{bmatrix}
\begin{bmatrix}
\hat{\boldsymbol d} \\ \hat{\boldsymbol n}
\end{bmatrix},
\\
\hat{\boldsymbol z}
&=
\begin{bmatrix}
\mathbf C_z & \mathbf 0\\
\mathbf 0 & - \alpha\mathbf K
\end{bmatrix}
\begin{bmatrix}
\hat{\boldsymbol q} \\ \hat{ \boldsymbol q}_e
\end{bmatrix}.
\end{align*}

\subsection{$\mathcal H_2$-norms: Uncontrolled flow}

The $\mathcal H_2$-norm for one channel half is
\begin{align}
\|\hat{\boldsymbol u}\|_2 =& \sqrt{\text{tr}(\mathbf{C}_z \mathbf Z \mathbf C_z^*)},
\end{align}
and at individual heights it is
\begin{align}
\|\hat{\boldsymbol u}(z)\|_2  =& \sqrt{\text{diag}(\mathbf{C} \mathbf Z \mathbf C^*)},
\end{align}
where $\mathbf{Z}$ is found by solving the following Lyapunov equation:
\begin{align}
\mathbf{A} \mathbf{Z} + \mathbf{Z}\mathbf{A}^* = -\mathbf{B}_d\mathbf{B}_d^*.
\end{align}
\subsection{$\mathcal H_2$-norms: Controlled flow}

The $\mathcal H_2$-norms for one channel half are
\begin{align}
\|\hat{\boldsymbol z}_{AE}\|_2 =& \sqrt{\text{tr}(\mathbf{C}_z \mathbf Y \mathbf C_z^*)},
\\
\|\hat{\boldsymbol z}_{ME}\|_2 =& \sqrt{\text{tr}(\mathbf{B}_d^* \mathbf X \mathbf B_d)},
\\
\|\hat{\boldsymbol z}_{IO}\|_2 =&  \sqrt{\text{tr}(\mathbf{C}_z \mathbf{Y C}_z^*) + \text{tr}( \mathbf C_m \mathbf{YXL} )} = \sqrt{\text{tr}(\mathbf{B}_d^* \mathbf{X B}_d) + \text{tr}(  \mathbf{KYX} \mathbf B_f)}.
\end{align}

The $\mathcal H_2$-norms at individual wall heights are
\begin{align}
\|\hat{\boldsymbol z}_{AE}(z)\|_2 &=  \sqrt{\text{diag}(\mathbf{C} \mathbf W_c \mathbf C^*)} = \sqrt{\text{diag}(\mathbf{C} \mathbf Y \mathbf C^*)},
\label{eqn:gammaOEsq}
\\
\|\hat{\boldsymbol z}_{ME}(z)\|_2  &= \sqrt{\text{diag}\left(\begin{bmatrix}
	\mathbf{C} \\ \mathbf{0}
	\end{bmatrix} \mathbf W_c \begin{bmatrix}
	\mathbf{C}  \\ \mathbf{0} 
	\end{bmatrix}^*\right)},
\\
\|\hat{\boldsymbol z}_{IO}(z)\|_2 &= \sqrt{\text{diag}\left(\begin{bmatrix}
	\mathbf{C} & \mathbf{0} \\ \mathbf{0} & \mathbf{0}
	\end{bmatrix} \mathbf W_c \begin{bmatrix}
	\mathbf{C} & \mathbf{0} \\ \mathbf{0} & \mathbf{0}
	\end{bmatrix}^*\right)},
\end{align}          
where $\mathbf{W}_c$ is the controllability Gramian that is found by solving the following Lyapunov equation (based on the LFT):
\begin{align}
\mathbf{A}_L \mathbf{W}_c + \mathbf{W}_c\mathbf{A}_L^* = -\mathbf{B}_L\mathbf{B}_L^*.
\end{align}

\subsection{$\mathcal H_2$-norms: Actuation force}
\label{J2:sec:ActForce}

The $\mathcal H_2$-norms for the actuator forces are

\begin{align}
\|\hat{\boldsymbol f}_{AE}\|_2 =& \sqrt{\text{tr}((\mathbf{C}_z \mathbf L \mathbf C_m) \mathbf W_c  (\mathbf C_z \mathbf L \mathbf C_m )^*)} = \sqrt{\text{tr}((\mathbf{C}_z \mathbf L \mathbf C_m) \mathbf Y  (\mathbf C_z \mathbf L \mathbf C_m )^*)},
\\
\|\hat{\boldsymbol f}_{ME}(z)\|_2 &= \sqrt{\text{tr}\left(\begin{bmatrix}
	\mathbf{0} \\ { \alpha \mathbf{K}}
	\end{bmatrix} \mathbf W_c \begin{bmatrix}
	\mathbf{0} \\ { \alpha \mathbf{K}}
	\end{bmatrix}^*\right)},
\\
\|\hat{\boldsymbol f}_{IO}(z)\|_2 &= \sqrt{\text{tr}\left(\begin{bmatrix}
	\mathbf{0} & \mathbf{0} \\ \mathbf{0} & { \alpha \mathbf{K}}
	\end{bmatrix} \mathbf W_c \begin{bmatrix}
	\mathbf{0} & \mathbf{0} \\ \mathbf{0} & { \alpha \mathbf{K}}
	\end{bmatrix}^*\right)}.
\end{align}

\bibliography{References.bib}

\end{document}